\documentclass[prb,aps,twocolumn,floatfix,amsmath,amssymb,superscriptaddress,tightenlines]{revtex4}
\usepackage{graphicx}% Include figure files
\usepackage{epstopdf}
\usepackage{amsmath}
\usepackage{amsfonts}
\usepackage{bm}
\usepackage{color}

\usepackage{amsthm}

\begin{document}
\title{Matrix Product Operators and Central Elements: Classical Description of a Quantum State}

\author{Matthew B. Hastings}
\affiliation{Duke University, Department of Physics, Durham, NC, 27708}
\affiliation{Microsoft Research, Station Q, CNSI Building, University of California, Santa Barbara, CA, 93106}

\newcommand{\be}{\begin{equation}}
\newcommand{\ee}{\end{equation}}

\newtheorem{lemma}{Lemma}
\newtheorem{definition}{Definition}
\newtheorem{theorem}{Theorem}
\newtheorem{corollary}{Corollary}

\begin{abstract} 
We study planar two-dimensional quantum systems on a lattice whose Hamiltonian is a sum of local commuting projectors of bounded range.  We consider whether or not such a system has a zero energy ground state.
To do this, we consider the problem as a one-dimensional problem, grouping all sites along a column into ``supersites"; using $C^*$-algebraic methods\cite{bv}, we can solve this problem if we can characterize the central elements of the interaction algebra on these supersite.
Unfortunately, these central elements may be very complex, making brute force impractical.  Instead, we show a characterization of these
elements in terms of matrix product operators with bounded bond dimension.  This bound can be interpreted as a bound on the number of particle types in lattice theories with bounded Hilbert space dimension on each site.
Topological order in this approach is related to the existence of certain central elements which cannot be ``broken" into smaller pieces without creating an end excitation.
Using this bound on bond dimension, we prove that several special cases of this problem are in NP, and we give part of a proof that the general case is in NP.
Further, we characterize central elements that appear in certain specific models, including toric code and Levin-Wen models, as either product operators in the Abelian case or matrix product operators with low bond dimension in the non-Abelian case; this matrix product operator representation may have practical application in engineering the complicated multi-spin interactions in the Levin-Wen models.
\end{abstract}
\maketitle

The subject of Hamiltonian complexity theory is devoted to the study of the computational complexity of various problems in quantum many-body physics. 
A general framework for this problem is as follows.
We consider a quantum system whose Hilbert space is the tensor product of $N$ different Hilbert spaces.  Each of these $N$ Hilbert spaces is referred to as the Hilbert space of a ``site".  We are interested in the case that each site
has Hilbert space dimension that is $poly(N)$ (indeed, in many practical settings it is $O(1)$).  The Hamiltonian will be a sum of at most $poly(N)$ terms, where each term in the Hamiltonian acting on at most $O(1)$ sites.  Often, a locality condition is imposed on these terms in the Hamiltonian: there is some given graph and each term in the Hamiltonian only acts on a set of sites which has small diameter with respect to the graph metric
(following physics terminology, the term ``site" refers both to a vertex of a graph as well as to a Hilbert space associated to that vertex, and also following physics terminology we refer to edges as ``bonds").  Hamiltonian complexity theory then addresses the complexity of answering various questions about this Hamiltonian; in this paper we address questions about ground state energy, but other problems including time dynamics and correlations have been considered by others.

One basic result is that determining whether or not a quantum many-body system has a ground state energy less than or equal to some given value is complete for the complexity class QMA\cite{kitaev}; this contrasts with the classical case where the analogous decision problem for the ground state energy of a classical system is complete for NP.  
One important area of research relates to the different promise gaps in these settings: the decision problem in the quantum context is phrased as deciding whether the energy is less than some value $E_0$ or greater than some value $E_1$ where the difference in energies is called the ``promise gap".  Roughly speaking, one can think of this as requiring one to approximate the ground state energy to an accuracy given by the promise gap.  Thus far, there is a big difference in the promise gaps in the quantum and classical settings.  The quantum result is known to hold only for promise gaps which are $1/poly(N)$ (though with these polynomially small gaps the result has been extended down even to one-dimension\cite{gkai}), while the classical result is known to hold for promise gaps which are of order the norm of the Hamiltonian itself as follows from the PCP theorem\cite{pcp1,pcp2,pcp3}.
While there has been much effort in trying to prove or disprove the quantum PCP conjecture\cite{qpcp1,qpcpme,qpcparad}, thus far the problem is open as to how large the promise gap can be made in the quantum setting to still have a QMA-complete problem.

Another important area of interest is the study of quantum Hamiltonians which are a sum of commuting projectors.  This set of quantum Hamiltonians lies in between the classical case and the quantum case. 
Thus far all QMA-completeness results for quantum Hamiltonians construct Hamiltonians which are not in this set of commuting projector Hamiltonians.
One nice feature of a commuting projector Hamiltonian is that all the eigenvalues are integers, so that the issue of a promise gap does not arise: that is, given that the Hamiltonian is a sum of commuting projectors one knows that if the energy is greater than some given $E_0$ then it is at least $\lfloor E_0 \rfloor+ 1$.

Now consider the difference between the commuting projector case and the classical case.
 In the classical case, not only do the individual terms in the Hamiltonian commute, they are all diagonal in a product basis and the ground state of the Hamiltonian is a product state.
In the case of commuting projectors, this may not be possible.  Indeed, there exist Hamiltonians whose ground state cannot be turned into a product state by a local quantum circuit with depth and range small compared to the system size.  
Prototypical examples of such Hamiltonians in two dimensions include the toric code\cite{tc} and Levin-Wen models\cite{lw}.  Such ground states are called {\it topologically ordered} states, and this definition that a state is topologically ordered if it cannot be turned into a product state by a local quantum circuit is a circuit definition of topological order (see \cite{circuit} among others).
On the other hand, other commuting projector Hamiltonians do have topologically trivial ground states, meaning that the ground state {\it can} be turned into a product state by a quantum circuit of small depth and range.  One example is when the Hamiltonian is a sum of terms acting only on two sites at a time\cite{bv} (see also \cite{qpcpme}).

Such topologically trivial states can act as a classical witness for the existence of a low energy eigenstate of the Hamiltonian.  
Given a state $\psi$ which is given by a circuit of depth $O(1)$ acting on some given product state, with each unitary in the circuit acting on at most $O(1)$ sites, one can compute the reduced density matrix of $\psi$ on any set of $O(1)$ sites in a time that is $poly(N)$; hence, one can compute the expectation value of each term in the Hamiltonian in a time that is $poly(N)$ and so one can compute the energy of the state $\psi$ in a time $poly(N)$.  So, if such a Hamiltonian has a topologically trivial ground state, there is a classical witness which can be check in polynomial time.

In this paper we show that for several special cases of commuting projector Hamiltonians in two dimensions as defined below, the problem of determining whether there is a zero energy ground state is in NP (the special cases are defined below; we only sketch part of a proof for the general case, leaving the rest for a future work). 
 Note that this includes Hamiltonians whose ground state is topologically ordered.  Hence, our classical witness cannot simply be a description of a trivial state using a quantum circuit.  Instead, our classical witness will include certain central elements of certain interaction algebras as discussed below.  These central elements may include ``string-like operators", as described below, characteristic of systems with topological order.  Thus the key step in this result is characterizing possible string-like operators which may appear.  These  operators are called ``string-like" because they may have support on a column of sites in the two-dimensional lattice (the number of such sites will be of order $L$, for a two-dimensional system of size $L$-by-$L$).  To specify an arbitrary operator with support on that many sites requires specifying exponentially many (in $L$) different matrix elements, and hence even if we only specify each matrix element by approximating it with polynomially many bits, the description of such an arbitrary operator would require exponential resources.  Hence, we will need to give a better characterization of these operators.

Note that it is known that the analogous problem for classical Hamiltonian is NP-complete\cite{barahona} and hence this quantum problem is also NP-complete as it contains the classical problem.  Thus, it should come as no surprise that we will not show how to find the witness efficiently.

We begin by defining the problem of two dimensional commuting local Hamiltonian, which we call 2DCLH, in section \ref{probdef}, and show how various different systems can be cast in the form of our problem definition.  Then, in section \ref{cstar} we review the $C^*$-algebraic approach to commuting Hamiltonians and the role of central elements in the interaction algebra.  We also define the ideas of ``propagation", ``breaking", and ``masking" which are used later.  We show how complicated central elements can arise in even classical models, but for such models we show how these elements can be simplified by a ``refinement" to a problem in a which the central element is a product of operators on a single sites, each such operator commuting with the Hamiltonian. 
In section \ref{example}, we discuss an example of central elements that appear in a simple toric code model with topological order.  We show how one cannot simplify the central elements in this case to a product of operators on single sites which commute with the Hamiltonian.  Indeed one cannot even ``break the string in two", as such a breaking would create endpoint excitations; this is defined later and is shown using central elements of a different interaction algebra which measure topological charge.
In section \ref{break}, we consider a restricted class of Hamiltonian where no such elements measuring topological charge do not exist, and show how conversely one can ``break" the string-like operators when these terms are absent and in section \ref{solv}, we show that such Hamiltonians have trivial ground states and hence there exists a witness to the existence of a zero energy ground state for these Hamiltonians.
Finally, in section \ref{solvgen}, we consider the problem of constructing a witness to the zero energy ground state for arbitrary instances of 2DCLH.  We prove that such a witness exists for another special case, and partially sketch the general case; the full proof of the general case will appear elsewhere.

\section{Problem Definition}
\label{probdef}
We now define the problem 2DCLH.
We consider a square lattice, with size $L$-by-$L$.  
We refer to one direction of this $L$-by-$L$ lattice as the vertical direction and one direction as the horizontal direction.  The sites will be described by horizontal and vertical coordinates each ranging from $0,...,L-1$.  

Let $N=L^2$ be the total number of sites.  On each site we have a Hilbert space with dimension $D$ at most polynomial in $N$.  The Hamiltonian that we consider is
\be
\label{projsum}
H=\sum_Z (1-P_Z)=\sum_Z Q_Z,
\ee
where $Q_Z=1-P_Z$ and the $P_Z$ are projection operators supported on a set of sites $Z$.
The sum is over sets $Z$ that consist of four sites around a plaquette.
We have
\be
[Q_Y,Q_Z]=0
\ee
for all $Y,Z$.

Note: throughout this paper, if we write an operator $O_A$, where $A$ is a set of sites, it is assumed that the support of $O_A$ is the set $A$ unless we specify otherwise.  Similarly, if we write $O_{ij}$ where $i,j$ are sites, then the support is the set $\{i,j\}$, or if we write $O_{ijk}$ where $i,j,k$ are sites then the support is the set $\{i,j,k\}$, and so on.  This convention applies also in the case of operators with additional superscripts or subscripts: for example, 
 we might define several operators $O_A^1, O_A^2, O_A^3, ...$ all supported on the set of sites $A$.  When we do this, we will usually introduce an index $\alpha$ taking a value from some finite set and write a given operator from that last $O_A^1,O_A^2,...$ as $O_A^\alpha$.

The problem 2DCLH is to determine whether or not $H$ has a zero energy ground state.
Throughout this paper, we gloss over details of real number arithmetic, but we make a few comments about this now. 
Suppose, first, that one wants to check that a Hamiltonian $H$ really is a sum of commuting projectors.
An operator $Q_Z$ is described by several real numbers, namely the matrix elements of the operator.  Using a polynomial number of bits to approximate each of these numbers, one can check that two projectors commute up to exponentially small error and one can check that each operator is a projector up to exponentially small error.
While we will describe our result later as showing that there is a classical witness for a state with zero energy given that the $P_Z$ are projectors and commute exactly, the witness will include some real numbers.  If we use a polynomial number of bits to specify these real numbers, what the verifier will be able to verify is that there is a state with energy less than some quantity which is smaller than an exponential of a polynomial in $N$ (rather than verifying the existence of a state with strictly zero energy).  We will omit the details of this, though, and instead describe a witness specified by some exact real numbers.
However,
given a promise that the Hamiltonian actually is a sum of commuting projectors, if one proves that there is a state with energy less than some exponentially small quantity, the existence of a zero energy state follows immediately, since the eigenvalues of the Hamiltonian must then all be non-negative integers.

Note that the restriction to projectors is not a serious restriction.  Suppose we have a Hamiltonian
$H=\sum_Z h_Z$ where $[h_Y,h_Z]=0$ for all $Y,Z$ but the $h_Z$ are not necessarily projectors.  Then, the ground state of $H$ can be chosen as an eigenvector of each $h_Z$ with eigenvalue $\lambda_Z$ with the ground state of $H$ having energy $\sum_Z \lambda_Z$.  Then, we can define a Hamiltonian which is a sum over commuting projectors by defining a new Hamiltonian $H'=\sum_Z (1-P_Z)$ where $P_Z$ projects onto the eigenspace of $h_Z$ with eigenvalue $\lambda_Z$.  Then $H'$ has a zero energy ground state if and only if there is a state which indeed is an eigenvector of every $h_Z$ with eigenvalue $\lambda_Z$.  So, the restriction to terms which are projectors in the Hamiltonian is not an important restriction.

While some other features of our problem may also seem like arbitrary choices (for example, a square lattice rather than hexagonal or triangular or other lattice, or the restriction to a planar problem rather than a problem on a cylinder or other topology), the problem definition is actually not that restrictive.  Consider any other two dimensional lattice with bounded range interactions on that lattice.  We can group several of the sites in the original lattice ``supersites" such that the resulting lattice of supersites {\it is} a square lattice and such that each interaction term in the original Hamiltonian acts only on sites in the same plaquette.  Adding all the interaction terms in a given plaquette together, we get a sum of commuting terms $H=\sum_Z h_Z$ as in the above paragraph.  Note that this grouping into supersites will entail some overhead, in that the dimension of each supersite will be larger than the original sites, but if the original interactions are bounded range then this will lead to only an $O(1)$ factor increase.

Also, the restriction to planar two dimensional systems is not so serious.  In subsection \ref{othergeom} we show how to cast Hamiltonians in cylindrical or spherical geometry as problems with planar interaction, by ``squashing the geometry flat".  This leads to some overhead, in that the dimension $D$ on each site of the resulting planar problem is larger than that of the original problem, but for cylinder and sphere (and other geometries with genus $O(1)$) this gives only an increase in the dimension to $D^{O(1)}$.

\section{$C^*$-algebraic Methods and Supersites}
\label{cstar}
We begin with a review of the method for finding trivial ground states of commuting projector Hamiltonians
which are a sum of two-site and one-site interaction terms.  
This is based on results of \cite{bv} and see also \cite{topotemp,qpcpme}.
We then discuss a grouping into supersites that we will use for the 2DCLH problem.

\subsection{$C^*$ Methods}
A key role is played by the concept of an interaction algebra,
which originates in Ref.~\onlinecite{intalg}.
\begin{definition}
Given an operator $O$ and a set $X$ we define the interaction algebra of $O$ on $X$ to be the algebra supported on $X$ generated by all operators of the form ${\rm tr}_{\overline X}(O Q_{\overline X})$ where the trace is over all sites in the complement of $X$ and $Q_{\overline X}$ is any operator supported on the complement of $X$.

More generally we can also define the interaction algebra of an algebra: given an algebra ${\cal A}$, we define
 the interaction algebra of ${\cal A}$ on $X$ to be the algebra supported on $X$ generated by all operators of the form ${\rm tr}_{\overline X}(O Q_{\overline X})$ where the trace is over all sites in the complement of $X$ and $Q_{\overline X}$ is any operator supported on the complement of $X$ and where $O$ is any operator in ${\cal A}$.

Finally, whenever a Hilbert space ${\cal H}$ decomposes as ${\cal H}={\cal H}_1 \otimes {\cal H}_2$, we can define the interaction algebra of an operator $O$ on a space ${\cal H}_1$ to be the algebra generated by all operators ${\rm tr}_{2}(O Q_2)$, where the trace is over space ${\cal H}_2$ and $Q_2$ is any operator supported on ${\cal H}_2$.  We can define the interaction algebra of an algebra on ${\cal H}_1$ similarly.
\end{definition}

We also define
\begin{definition}
Given an algebra, we say that the projectors $P^\alpha$ are the {\bf minimal} projectors which generate the center of the algebra if $P^\alpha$ are projectors for each $\alpha$ that generate the center of the algebra so that $P^\alpha P^\beta=0$ for $\alpha \neq \beta$.
\end{definition}

We write 
a two-body Hamiltonian as
\be
\label{twobody}
H=\sum_{<i,j>} H_{i,j}+\sum_{i} H_{i,i},
\ee
where $H_{i,j}$ acts only on sites $i,j$ and $H_{i,i}$ acts only on site $i$.  The $H_{i,j}$ and $H_{i,i}$ need not be projectors.  
The notation $\sum_{<i,j>}$ indicates a summation over all pairs of sites $i$ and $j$ that are neighbors in some given graph; in the next subsection, we will talk about a grouping that reduces certain problems to the case where the graph is a line, while in this subsection the graph is not specified.

Suppose all the various terms $H_{i,j}$ and $H_i$ commute with
each other.  Consider any site $i$.  Let ${\cal A}^{ij}$ be the interaction algebra of $H_{i,j}$ on $i$.  The interaction algebras ${\cal A}^{ij}, {\cal A}^{ik}$ commute for $j \neq k$ and also these algebras commute with $H_{i,i}$.

Let ${\cal H}_i$ denote the Hilbert space on site $i$.  We assume that ${\cal H}_i$ is finite dimensional.
Then, it is a fact from $C^*$-algebra that
we can decompose ${\cal H}_i$ into a direct sum of Hilbert spaces ${\cal H}_i^{\alpha(i)}$,
\be
\label{decompsum}
{\cal H}_i=\bigoplus_{\alpha(i)} {\cal H}_i^{\alpha(i)},
\ee
where $\alpha(i)$ is some discrete index,
 and then further decompose each such Hilbert space ${\cal H}_i^{\alpha(i)}$ into a tensor product of spaces ${\cal H}_{i\rightarrow j}^{\alpha(i)}$ (where the product ranges over $j$ that neighbor $i$) tensor producted with space ${\cal H}_{i, i}^{\alpha(i)}$ so that
\begin{eqnarray}
\label{decomp1}
{\cal H}_i& = &\bigoplus_{\alpha(i)} {\cal H}_i^{\alpha(i)}
= \bigoplus_{\alpha(i)} \Bigl( {\cal H}_{i,i}^{\alpha(i)} \otimes \bigotimes_{<j,i>} {\cal H}_{i\rightarrow j}^{\alpha(i)} \Bigr),
\end{eqnarray}
where the product is over $j$ that neighbor $i$,
such that each operator $H_{i,j}$ can be decomposed as
\be
\label{decomp2}
H_{i,j}=\sum_{\alpha(i),\alpha(j)} P_i^{\alpha(i)} P_j^{\alpha(j)} H_{i,j}^{\alpha(i),\alpha(j)},
\ee
where $P_i^{\alpha(i)}$ is the operator on ${\cal H}_i$ which projects onto ${\cal H}_i^{\alpha(i)}$ and $H_{i,j}^{\alpha(i),\alpha(j)}$ acts
on the subspace of ${\cal H}_i^{\alpha(i)} \otimes {\cal H}_j^{\alpha(j)}$ given by ${\cal H}_{i\rightarrow j}^{\alpha(i)} \otimes {\cal H}_{j \rightarrow i}^{\alpha(j)}$ and such that
$H_{i,i}$ can be decomposed as
\be
\sum_{\alpha(i)} P_i^{\alpha(i)} H_{i,i}^{\alpha(i)}
\ee
where $H_{i,i}^{\alpha(i)}$ acts only on ${\cal H}_{i,i}^{\alpha(i)}$.

The operators $P_i^{\alpha(i)}$ commute with each other and commute with the Hamiltonian for any $i$ and any $\alpha$.  
Thus, we can choose a basis for the space of ground states of $H$ such that every vector in this basis
is an eigenvector of all of these operators $P_i^{\alpha(i)}$.
The $P_i^{\alpha(i)}$ can be taken to be the minimal projectors which generate the center of the algebra generated by the interaction algebras on $i$ of $H_{i,j}$ for all $j$.
We find it useful to define an ``effective classical Hamiltonian":
\begin{definition}
Given a Hamiltonian which is a sum of commuting terms, each acting on at most two sites, we define the {\bf effective classical Hamiltonian} $H^{eff}$ as follows.  We will give two definitions, one as a function from a discrete set to the real numbers, while the second definition will be as an operator.  One can straightforwardly translate between these definitions and it should be clear in context which definition we mean.

Let $P_i^{\alpha(i)}$ be the projectors onto central elements defined above.
Let
\be
H^{eff}(\alpha(1),\alpha(2),...,\alpha(N))={\rm min}_{\psi, |\psi|=1, P_i^{\alpha(i)}\psi=\psi} \langle \psi,H \psi \rangle.
\ee
That is, $H^{eff}$ is the minimum over all states $\psi$, with norm $|\psi|=1$ and such that $P_i^{\alpha(i)}\psi=\psi$ of $\langle \psi,H \psi \rangle$.

Similarly, we can define a quantum operator
\begin{eqnarray}
&& H^{eff}
\\ \nonumber
&=&\sum_{\alpha(1),\alpha(2),...,\alpha(N)} \Bigl( \prod\limits_{i=1}^N P_i^{\alpha(i)} \Bigr) H^{eff}(\alpha(1),\alpha(2),...,\alpha(N)).
\end{eqnarray}
\end{definition}

We noted that there is a basis of ground states such that all basis vectors are eigenvectors of all the operators $P_i^{\alpha(i)}$, so there is a ground
state $\psi$  such that $P_i^{\alpha(i)} \psi=\psi$ for all $i$ for some given choice of the indices $\alpha(i)$.   
Given the decomposition (\ref{decomp2}), we can construct such a state $\psi$ which has simple entanglement properties: 
this state $\psi$ is a product of states $\psi_{i,j}$ and states $\psi_i$, where $\psi_{i,j}$ is in
the space ${\cal H}_{i \rightarrow j}^{\alpha(i)}\otimes {\cal H}_{j \rightarrow i}^{\alpha(j)}$ and$\psi_{i}$ is in ${\cal H}_{i,i}^{\alpha(i)}$.
As shown in \cite{qpcpme}, such a state can be generated by a quantum circuit of small range and depth acting on a product state.
Note that given a Hamiltonian $H$, if we can determine a choice of $\alpha(1),...,\alpha(N)$ which minimzes $H^{eff}(\alpha(1),...,\alpha(N)$ then we can construct the states $\psi_{i,j}$ in a time polynomial in $N$ (again, we gloss over the details of the real number arithmetic required to do this) since each such state can be obtained by diagonalizing a Hamiltonian acting only on the space ${\cal H}_{i \rightarrow j}^{\alpha(i)}\otimes {\cal H}_{j \rightarrow i}^{\alpha(j)}$ which has dimension at most polynomial in $N$.

\subsection{Grouping into Supersites}
\begin{figure}
\includegraphics[width=1.9in]{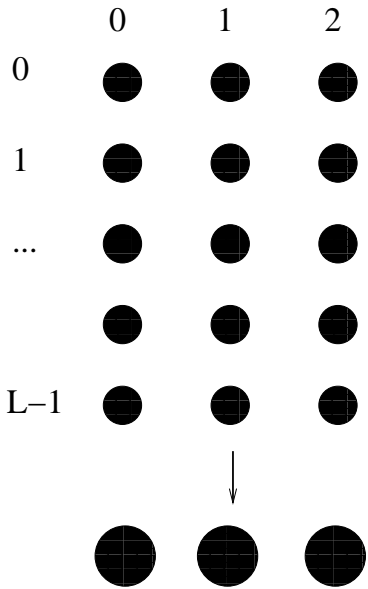}
\caption{Illustration of grouping.  Smalled filled circles are sites.  Vertical coordinates are shown ranging from $0$ to $L-1$.  $3$ of the columns are shown with horizontal coordinates $0,1,2$.  The figure below the arrow represents all sites in a single column grouped into a supersites with dimension $D^L$.}
\label{figgroup}
\end{figure}

The discussion above applies to Hamiltonians which are a sum of two-body terms.
However, our Hamiltonian Eq.~(\ref{projsum}) is not a sum of two-body terms since each projector may act on four sites in a plaquette.
See Fig.~\ref{figgroup}.
We group all sites in a vertical column into a single supersite.  We label the columns by numbers $C$.  In an abuse of notation, we use $C$ both as a number between $0$ and $L-1$ labelling the horizontal coordinate of these sites in the given column and we also use $C$ as a set (so that a site $i$ is in the set $C$ if and only if $i$ has horizontal coordinate $C$).
Then, the Hamiltonian Eq.~(\ref{projsum}) is a sum of terms each acting on a pair of supersites.  Thus, we have succeeded in turning the 2DCLH  problem into a problem of two-body interactions.  However, this comes at a large cost: the dimension of the Hilbert space on each supersite is $D^L$, so it is exponentially large.  Writing down an arbitrary operator on a single supersite requires exponential resources.  The main work of this paper will be to show how to deal with this, by better characterizing the central elements of the interaction algebra that can appear in this problem.

We need some definitions now.  Recall that given an algebra ${\cal A}$, the center of the algebra is the set of elements of the algebra that commute with every other element of the algebra.  These elements are also called central elements.  
\begin{definition}
 The projectors $Q_Z$ generate a subalgebra of the algebra of all possible operators on the $D^N$ dimensional Hilbert space.  We call this subalgebra the {\bf term algebra}.

Given a set $S$, we say that the {\bf term algebra of} $S$ is the algebra generated by all $Q_Z$ with $Z\subset S$.  Given sets $S,B$, we define the {\bf interaction algebra of $S$ on $B$} to be the interaction algebra of the term algebra of $S$ on $B$.
\end{definition}

We define that
\begin{definition}
A projector $P$ is {\bf consistent} with a Hamiltonian $H$ written in the form of Eq.~(\ref{projsum}) if
there is a zero energy ground state of $H$ that is in the range of $P$.
Note that for a projector to be consistent with a Hamiltonian, that Hamiltonian must have at least one zero energy state.

A projector $P'$ is a {\bf refinement} of $P$ if $P'\leq P$.
$P'$ is a {\bf consistent refinement} of $P$ if $P'$ is consistent with the Hamiltonian $H$ and $P'$ is a refinement of $P$.
\end{definition}

Finally, we define that
\begin{definition}
An operator $O_X$ supported on a set $X$ which commutes with the term algebra is {\bf breakable} into sets $Y_1,Y_2,...$ if we can write
\be
\label{breakabledef}
O_X=\sum_{\alpha} \Bigl( \prod_a O_{Y_a}^\alpha \Bigr)
\ee
where the sum is over $\alpha$ in some finite set (when we write an operator as a sum of product of operators with a sum of this form we will suppress the set from which $\alpha$ is chosen when writing the sum, as it should be clear from context) and
for some choice of operators $O_{Y_a}^\alpha$ such that $O_{Y_a}^\alpha$
is supported on set $Y_a$ and $O_{Y_a}^\alpha$ commutes with the term algebra for all $a,\alpha$.
The operators $O_{Y_a}^\alpha$ need not commute with $O_{Y_b}^\alpha$; if they do not commute, then we require that there be some choice of the ordering in
Eq.~(\ref{breakabledef}) for which the equation holds and different orderings can be chosen for different $\alpha$ in the sum.
\end{definition}

The reason for this last definition is that we will be interested in projectors that are central elements of the interaction algebra that are {\it not} breakable into sets of small diameter or which cannot be refined into projectors which are breakable on sets of small diameter.  Let us consider first a case with breakable projectors to illustrate what can happen when the projectors which appear as central elements of the interaction algebra are breakable.
Consider as an extreme case a classical model in two dimensions with nearest neighbor interactions, where by classical we mean that every interaction term is diagonal in some product basis.  Let this product basis be the product of the basis on each site with states labeled $|1\rangle,|2\rangle,...|D\rangle$.
Then, the interaction algebra of $h_{C-1,C}$ on column $C$ includes central elements which are also diagonal in this product basis.  However, assuming that $H$ has a zero energy ground state then in this case all of these central elements which are consistent with $H$ have a consistent refinement to a projector $P$ which is a product of projectors on single sites in the column with each such single site projector having rank $1$, so that
\be
P_C=\prod_{i \in C} P_i 
\ee 
 To see this, simply take any zero energy ground state of $H$ which is a product state (such a state exists because the Hamiltonian is classical) and let $P_i$ be the reduced density matrix of that state on site $i$.  Each of these projectors $P_C$ commutes with the term algebra.
When the central elements are breakable in this fashion, we can use the projectors on small sets as a classical witness for the existence of a zero energy state.
However, we will see in the next section that in certain systems with topological order there exist central elements which are not breakable into small sets as discussed in subsection \ref{unbreak}.

\subsection{Propagation}
Finally, we describe an iterative method of determining whether a given instance of 2DCLH has a zero energy ground state.  In order to verify that there is a zero energy ground state, we wish to verify that
${\rm tr}(\prod_Z P_Z)\geq 1$ as this trace counts the number of zero energy states.
We can compute this trace using the following iterative procedure.  
Define
\be
P_{C,C+1}=\prod_{Z \subset C \cup C+1} P_Z,
\ee
where the product is over sets $Z$ contained in the union of column $C$ with column $C+1$.
For use later, define
\be
P_{C,C'} = \prod_{Z\subset C \cup C+1 \cup C+1 \cup ... \cup C'} P_Z,
\ee
for any pair of columns $C,C'$ with $C<C'$.

Define $\rho_C$ by
\be
\label{rho0I}
\rho_0=I
\ee
and for $C\geq 0$ define
\be
\label{propeq1}
\rho_{C+1}={\rm tr}_C \Bigl(\rho_C P_{C,C+1} \Bigr),
\ee
where the trace is over all sites in column $C$.  Then,
\be
{\rm tr}\Bigl( \prod_Z P_Z \Bigr)={\rm tr}_{L-1}\Bigl(\rho_{L-1}\Bigr).
\ee

The operators $\rho_C$ can be written as matrix product operators\cite{mpo,mpo2,mpoexp}.
However, the bond dimension required to write a given operator $\rho_C$ may grow exponentially with $C$, so the above method may take exponential time.  To show that the problem is in NP, later we will construct a classical witness which will help reduce the bond dimension of the matrix product operators that we need to consider.
Note that each $\rho_C$ is in the interaction algebra of $C-1 \cup C$  on column $C$ (recall that in our terminology this interaction algebra is the interaction algebra on column $C$ of the algebra generated by the $Q_Z$ supported on the union of column $C-1$ with column $C$).
Things can be simplified to some extent, since we only care about the center of this algebra.  Let $P^\alpha_C$ be the minimal projectors which generate the center of this algebra.  That is, the center of this interaction algebra is the algebra generated by the $P^\alpha_C$ and we choose these projectors such that ${\rm tr}(P^\alpha_C P^\beta_C)=\delta_{\alpha,\beta}$.  Then defining
\be
\label{phiCdef}
\phi_C=\sum_\alpha \frac{P^\alpha_C {\rm tr}(P^\alpha_C \rho_C)}{{\rm tr}_C(P^\alpha_C)},
\ee
we claim that Eq.~(\ref{propeq1}) is the same as
\be
\label{propeq2}
\rho_{C+1}={\rm tr}_C \Bigl(\phi_C P_{C,C+1} \Bigr),
\ee
and we show this below.
We say that $\rho_{C+1}$ is the {\bf propagation} of $\rho_C$ from $C$ to $C+1$.  The propagation of $\rho_C$ from column $C$ to $C'$ for $C'>C$ is given by ${\rm tr}_{C,C+1,...,C'-1} \Bigl(\rho_C P_{C,C'}\Bigr)$.  Note that this is equivalent to propagating $\rho_C$ from $C$ to $C+1$, then propagating the result to $C+2$, and so on, until reaching column $C'$.
(Note also that a similar idea of propagation was used in \cite{mprodinf} as a heuristic method of finding ground states of classical Hamiltonians using a heuristic method to reduce the bond dimension.)

We now show that Eq.~(\ref{propeq2}) and Eq.~(\ref{propeq1}) give the same result for $\rho_{C+1}$.  
The proof we now give will also show that if we had instead chosen the $P^\alpha_C$ to be the minimal projectors which generate the center of the term algebra then again Eqs.~(\ref{propeq2}) and Eq.~(\ref{propeq1}) give the same result for $\rho_{C+1}$.  
Let ${\cal H}_C$ be the Hilbert space on column $C$.  This Hilbert space decomposes as a direct sum of Hilbert spaces ${\cal H}_C^\alpha$, with ${\cal H}_C^\alpha$ being the range of $P^\alpha_C$, and each ${\cal H}_C^\alpha$ decomposes as a product ${\cal H}_{C \rightarrow C-1}^\alpha$ and ${\cal H}_{C \rightarrow C+1}^\alpha$ such that the interaction algebra of $C-1 \cup C$ on $C$ acts only on ${\cal H}_{C \rightarrow C-1}^\alpha$ and the interaction algebra of $C \cup C+1$ on $C$ acts only on ${\cal H}_{C \rightarrow C+1}^\alpha$.
Previously we used this decomposition in the case of a set of minimal projectors which generate the center of the interaction algebra of the term algebra on column $C$, but it also holds here.
Then, by Eq.~(\ref{propeq1}) we have
\begin{eqnarray}
\label{asintextabove}
\rho_{C+1}&=&{\rm tr}_C (\rho_C P_{C,C+1} ) \\ \nonumber
&=& \sum_\alpha {\rm tr}_C (P^\alpha_C \rho_C  P_{C,C+1} ).
\end{eqnarray}
However, since $\rho_C$ acts only on ${\cal H}_{C \rightarrow C-1}^\alpha$ and $P_{C,C+1}$ acts only on ${\cal H}_{C \rightarrow C+1}^\alpha$,
\begin{eqnarray}
&&{\rm tr}_C (P^\alpha_C \rho_C P_{C,C+1})
\\ \nonumber
& = & {\rm tr}_{{\cal H}_{C \rightarrow C-1}^\alpha}(P^\alpha_C \rho_C) {\rm tr}_{{\cal H}_{C \rightarrow C+1}^\alpha}(P^\alpha_C P_{C,C+1}) \\ \nonumber
&=&
\frac{  {\rm tr}_{C}(P^\alpha_C \rho_C)    } {   {\rm tr}_{{\cal H}_{C \rightarrow C+1}^\alpha}(P^\alpha_C) }
\frac{  {\rm tr}_{C}(P^\alpha_C P_{C,C+1})    } {   {\rm tr}_{{\cal H}_{C \rightarrow C-1}^\alpha}(P^\alpha_C) }
\\ \nonumber
&=&
\frac{ {\rm tr}_C ( P^\alpha_C \rho_C) {\rm tr}_C(P^\alpha_C P_{C,C+1}) }{{\rm tr}_C(P^\alpha_C)},
\end{eqnarray}
and so Eq.~(\ref{propeq2}) follows.  In the above equation, traces such as ${\rm tr}_{{\cal H}_{C \rightarrow C+1}^\alpha}(...)$ or ${\rm tr}_{{\cal H}_{C \rightarrow C-1}^\alpha}(...)$ represent partial traces over the appropriate space; in every case where we write such a trace, the operator inside the trace is nonvanishing only on the subspace ${\cal H}_C^\alpha$.

We now define ``masking":
\begin{definition}
We say that an operator $O_C$ supported on a column $C$ {\bf masks} a given operator $O_{C-1}$ supported on column $C-1$ if
\begin{eqnarray}
\label{maskdef}
&& {\rm tr}_{C-1 \cup C} \Bigl(O_{C-1} P_{C-1,C} O_{C} P_{C,C+1} \Bigr)
\\ \nonumber
&=&
x {\rm tr}_C \Bigl(O_C P_{C,C+1} \Bigr)
\end{eqnarray}
for some scalar $x$.

We say that $O_C$ is a mask if it masks all operators $O_{C-1}$ supported on column $C-1$ (note that the constant $x$ may depend upon the given operator $O_{C-1}$ chosen).
\end{definition}
Note that the constant $x$ in Eq.~(\ref{maskdef}) may equal zero.

We claim that every minimal central element of the interaction algebra of $C-1 \cup C$ on column $C$ is a mask and also that
every minimal central element of the interaction algebra of $C-1 \cup C \cup C+1$ or $C \cup C+1$ on column $C$ is a mask.
Let us show the first of these claims (the other two claims are shown similarly).
As we did above Eq.~(\ref{asintextabove}), we decompose ${\cal H}_C$ as a direct sum of Hilbert spaces ${\cal H}_C^\alpha$, with ${\cal H}_C^\alpha$ being the range of $P^\alpha_C$, 
and decompose each ${\cal H}_C^\alpha$ into a product ${\cal H}_{C \rightarrow C-1}^\alpha$ and ${\cal H}_{C \rightarrow C+1}^\alpha$.
Then, for any $\alpha$, consider the trace
${\rm tr}_{C-1 \cup C}(O_{C-1} P_{C-1,C} P^\alpha_C P_{C,C+1})$.  Let ${\cal H}^\alpha_L={\cal H}_{C-1} \otimes {\cal H}_{C \rightarrow C-1}^\alpha$.  The operators $O_{C-1}$ and $P_{C-1,C}$ act only on ${\cal H}_L^\alpha$ and the operator $P_{C,C+1}$ acts only on ${\cal H}_{C \rightarrow C+1}^\alpha$ so we can write this trace as
\begin{eqnarray} \nonumber
&&{\rm tr}_{{\cal H}_{L}^\alpha}(O_{C-1} P_{C-1,C} P^\alpha_C) 
 {\rm tr}_{{\cal H}_{C \rightarrow C+1}^\alpha}(P^\alpha_C P_{C,C+1} ) \\ \nonumber.
&=&{\rm tr}_{{\cal H}_{L}^\alpha} (O_{C-1} P_{C-1,C} P^\alpha_C)
 \frac{ {\rm tr}_C( P^\alpha_C P_{C,C+1}) } { {\rm tr}_{{\cal H}_{C \rightarrow C-1}^\alpha}( P^\alpha_C) }.
\end{eqnarray}
The denominator and the first term of the last line of the above equation are scalars, so Eq.~(\ref{maskdef}) follows.

Masking can be used to construct a witness for the existence of a low energy state as follows.
\begin{lemma}
\label{maskuse}
Suppose that for each column $C$, there exists some projector $P_C$ which is a mask (these projectors $P_C$ should not be confused with the projectors $P_Z$ used to define $H$; it should be clear by the subscript which projector is meant).  Suppose that $\prod_C P_C$ is consistent with $H$
and suppose that
\be
\label{commmask}
[P_C,P_{C-1,C}]=0
\ee
for all $C$ with $1\leq C \leq L-1$.
Suppose finally that each such $P_C$ is represented as a matrix product operator with bond dimension that is $poly(N)$.  Then, given such projectors written as matrix product operators, one can verify the existence of a zero energy state in a time polynomial in $N$.

Further, even if such projectors $P_C$ are given only for some subset of the columns $C$, with the projectors being given for the columns in the sequence $C_1,C_2,...,C_n$ for some $n$ with $1\leq C_1<C_2<...<C_n \leq L$, such that $C_{i+1}-C_i\leq O(1)$ for all $i$ and such that $L-C_n\leq O(1)$ and $C_1-1\leq O(1)$, one can still verify the existence of a zero energy state.
\begin{proof}
As mentioned before, we gloss over details of real number arithmetic, assuming all calculations are done to infinite precision.

We only describe the case in which projectors are given for every column $C$ in detail; the second case can be reduced to the first case by defining a new problems where columns $1,...,C_1$ are grouped into a single column (that is, all sites in those columns with a given vertical coordinate are grouped into a single site in the new problem with the same vertical coordinate), columns $C_1+1,...,C_2$ are grouped into a second column, and so on.  Since $C_{i+1}-C_i\leq O(1)$, the Hilbert space dimension of the new problem is $D^{O(1)}$.

The verifier first checks that Eq.~(\ref{commmask}) holds.  Since the $P_C$ are given as matrix product operators, this can be done efficiently.

Since $\prod_C P_C$ is consistent with $H$, ${\rm tr}(\prod_C P_C \prod_Z P_Z)>0$.  Conversely, given that ${\rm tr}(\prod_C P_C \prod_Z P_Z)>0$, the existence of a zero energy state follows.  So, the verifier just needs to check that ${\rm tr}(\prod_C P_C \prod_Z P_Z)>0$.
However,
\begin{eqnarray}
{\rm tr}(\prod_C P_C \prod_Z P_Z)&=&{\rm tr}(\prod_C P_C \prod_{C=1}^{L-1} P_{C,C+1})
\\ \nonumber
&=& {\rm tr}(P_1 P_{1,2} P_2 P_{2,3} ... P_{L-1} P_{L-1,L} P_L),
\end{eqnarray}
where the last line follows by Eq.~(\ref{commmask}).

For each column $C$ with $1\leq C \leq L-2$, the verifier computes the propagation of $P_C$ from $C$ to $C+1$, calling the result $\rho_{C+1}$.  The result is a matrix product operator with bond dimension $poly(N)$, and this calculation can be done efficiently (note that the bond dimension of the resulting operator is at most a $D^{O(1)}$ factor bigger than the bond dimension of $P_C$.  The verifier then propagates $\rho_{C+1} P_{C+1}$ to $C+2$ and compares to $\rho_{C+2}$.  The verifier checks that the propagation of $\rho_{C+1} P_{C+1}$ to $C+2$ is indeed equal to a constant times $\rho_{C+2}$, as in Eq.~(\ref{maskdef}).  Call the constant for which Eq.~(\ref{maskdef}) holds $c_{C+1}$.  The verifier also checks that the constant $c_{C+1}$ is positive.
Since
\be
\label{checkpos}
{\rm tr}(P_1 P_{1,2} P_2 P_{2,3} ... P_{L-1} P_{L-1,L} P_L)=\prod_{C=2}^{L-1} c_C {\rm tr}(P_{L-1} P_{L-1,L}),
\ee
checking positivity of each $c_C$ and of the trace on the right-hand side of Eq.~(\ref{checkpos}) is equivalent to checking positivity of ${\rm tr}(\prod_C P_C \prod_Z P_Z)$.
\end{proof}
\end{lemma}

\section{Central Elements in Toric Code and Other Models}
\label{example}
We now describe a model with central elements that cannot be broken into certain smaller sets.  We begin with a description of a toric code model on a square lattice and describe the central elements and effective classical Hamiltonian.  We then show the unbreakability.  Next, we consider some other geometries.  Finally, we discuss the central elements that can appear in Levin-Wen models among others, where the central elements are matrix product operators.

Consider the toric code Hamiltonian\cite{tc}: we have a square lattice with spin-$1/2$ degrees of freedom on each {\it site} of the lattice (often the degrees of freedom are placed on the bonds in the literature but we place them on the sites here).  The Hamiltonian is a sum of commuting terms.  Color the plaquettes of the square lattice alternately as light or dark plaquettes, like a checkerboard.  The Hamiltonian is
\be
H=-\sum_{P\in light} \prod_{i \in P} \sigma^z_i - \sum_{P \in dark} \prod_{i \in P} \sigma^x_i,
\ee
where the first sum is over light plaquettes $P$ and the product is over sites $i$ in the plaquette and the second sum is over dark plaquettes and the operators $\sigma^x_i,\sigma^z_i$ are Pauli spin operators on site $i$.
Note that here we choose to consider Hamiltonians which are not a sum of commuting projectors in order to keep our notation closer to the literature; recall the discussion of this in section \ref{probdef}.

We will discuss this Hamiltonian on various geometries, namely square, cylinder, torus, and sphere geometries.  We begin with the square geometry, as in section \ref{probdef}.  In the square geometry case, we will add some additional terms to the Hamiltonian on the boundaries: on each pair of neighboring sites $i,j$ on either the top or bottom edge of the system (these two sites will be connected by a horizontal edge), we add the term $-\sigma^x_i \sigma^x_j$ if those two sites are in a light plaquette.
Now, group all sites along a single vertical column $C$ into one ``supersite" as described above.  Define $O^x_C$ to be the product of $\sigma^x_i$ over all sites $i$ in a given column $C$:
\be
O^x_C=\prod_{i\in C} \sigma^x_i.
\ee\
The operator $O^x_C$ is a
central element of the interaction algebra of $C \cup C-1 $ on $C$ and also of the interaction algebra of $C \cup C+1$ on $C$ for all $C$.
Further, the effective classical Hamiltonian
\be
\label{effclasstc}
H^{eff}=\sum_{C=0}^{L-1} h^{eff}_{C,C+1},
\ee
involves these central elements,
where
\be
h^{eff}_{C,{C+1}}=-O^x_C O^x_{C+1}.
\ee

This square lattice geometry shows the feature we have mentioned: the existence of central elements in the interaction algebra which cannot be broken into certain smaller sets, as we will show in the next subsection.  The physical reason for this unbreakability is that breaking the operator will necessarily lead to the creation of an anyon excitation.

\begin{figure}
\includegraphics[width=1.9in]{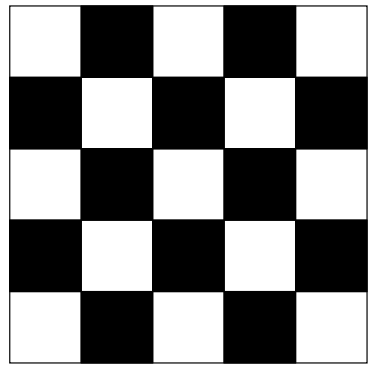}
\caption{Lattice for the toric code with $L=6$.  Light and dark plaquettes are indicated, with light plaquettes in the four corners of the system.  Interactions are as described in the text.}
\label{figtc}
\end{figure}

 However, this square lattice geometry also lacks one interesting feature: there is no term in the effective classical Hamiltonian on either the first or last supersite which would constrain the central element $O^x_C$ on that super-site to have any particular value.  That is, this Hamiltonian has two degenerate ground states corresponding to different values of $O^x_C$.  We can add this interesting feature to the Hamiltonian by adding additional interaction terms on the left and right edge of the system.  On each pair of neighboring sites $i,j$ on either the left or right edge of the system (these two sites will be connected by a vertical edge), we add the term $-\sigma^x_i \sigma^x_j$ if those two sites are in a light plaquette.  Further, we choose the geometry so that the four corners of the system are all light plaquettes (this means that we choose $L$ to be even).  Note that the corner sites now have two such added interaction terms, $-\sigma^x_i\sigma^x_j$, one coupling a corner site $i$ to a site $j$ neighboring it by a vertical bond and one coupling it to a site $j$ neighboring it by a horizontal bond.
See Fig.~\ref{figtc}.

By adding these terms to the Hamiltonian, we have now the effective classical Hamiltonian
\be
\label{effclasstc2}
H^{eff}=\sum_{C=0}^{L-1} h^{eff}_{C,{C+1}}-O^x_0-O^x_{L-1},
\ee 
which now has a unique ground state that minimizes every term in the Hamiltonian separately.
We could also have considered a slightly different Hamiltonian, by picking one of the vertical edges on the left side of the sample and changing the sign of interaction term so that we add $+\sigma^x_i \sigma^x_j$ rather than $-\sigma^x_i \sigma^x_j$.  In this case, the effective classical Hamiltonian is 
\be
H^{eff}=\sum_{C=0}^{L-1} h^{eff}_{C,C+1}+O^x_0-O^x_{L-1},
\ee
which has no ground states that minimize every term in the Hamiltonian.

\subsection{Unbreakability of Central Element}
\label{unbreak}
Recall that we each site has a horizontal and a vertical coordinate, each coordinate ranging from $0,...,L-1$.  A column $C$ contains all sites with a given horizontal coordinate.  

Consider the toric code Hamiltonian on the square lattice with these added operators  $-\sigma^x_i \sigma^x_j$  on the boundary.
We now show that
\begin{lemma}
\label{tcunbreak}
Suppose that $2\leq C \leq L-3$.
The central element $O^x_C$ is not breakable into any sets $Y_1,Y_2,...$ which have the properties that 
\begin{itemize}
\item[1] For all $a$ the set $Y_a$ does not contain any sites with horizontal coordinate $0$ or $1$ or horizontal coordinate $L-2,L-1$.
\item[2] At least one of the following two properties hold:
\subitem[2a]$Y_a$ does not contain any sites with vertical coordinate $0$.
\subitem[2b] $Y_a$ does not contain any sites with vertical coordinate $L-1$.
\end{itemize}

Remark: requirement (2) above can be roughly stated as $Y_a$ can be close to the top edge or to the bottom edge, but not to both edges while requirement (1) can be roughly stated as $Y_a$ cannot be close to either the left or right edge.

\begin{proof} Suppose instead $O^x_C$ is breakable.
Let $M$ be the set of sites which have horizontal coordinate $0$ or $L-1$, or have vertical coordinate $0$ or $L-1$, or both.  That is, $M$ is the set of sites on the boundary of the system.
Define the operator $T$ to be the product of $\sigma^z_i$ over all sites $i$ in $M$. 

The operator $T$ is in the algebra generated by the $Q_Z$, because the product of $Q_Z$ for all light plaquettes is equal to $T$.  Indeed, this operator $T$ is a central element of the interaction algebra of the term algebra on $M$.
Since each operator $O^\alpha_{Y_a}$ commutes with each $Q_Z$,
\be
\label{commT}
[O^\alpha_{Y_a},T]=0.
\ee
Note further that $[O^x_C,T]=0$.

Define $T^t$ to be the product of $\sigma^z_i$ over all $i$ in $M$ which have vertical coordinate less than or equal to $L/2$ and let $T^b$ be the product of $\sigma^z_i$ over all $i$ in $M$ which have vertical coordinate greater than $L/2$, so that
\be
T=T^t T^b.
\ee
Note that
\be
\{ O^x_C,T^t\}=\{O^x_C,T^b\}=0.
\ee

For each $a$, at least one of property 2a or 2b holds.  Suppose 2b holds.  Then, $[O^\alpha_{Y_a},T^b]=0$ since the operators $O^\alpha_{Y_a}$ and $T^b$ have disjoint support.
By Eq.~(\ref{commT}) we have
\begin{eqnarray}
O^\alpha_{Y_a} T^t T^b &=& T^t T^b O^\alpha_{Y_a} \\ \nonumber
&=& T^t O^\alpha_{Y_a} T^b.
\end{eqnarray}
Since $T^b$ is an invertible operator, we can multiply the right-hand side of the above equation by $(T^b)^{-1}$ to obtain
\be
[O^\alpha_{Y_a},T^t]=0.
\ee
Thus, $O^\alpha_{Y_a}$ commutes with both $T^t$ and $T^b$.  We can similarly show that $O^\alpha_{Y_a}$ commutes with both $T^t$ and $T^b$ if 2a holds.

So, the product over all $a$ of $O^\alpha_{Y_a}$ commutes with $T^t$ and $T^b$ for each $\alpha$.  However, $O^x_C$ anti-commutes with both $T^t$ and $T^b$.  Therefore, $O^x_C$ cannot be equal to $\sum_\alpha \prod_a O^\alpha_{Y_a}$.
\end{proof}
\end{lemma}

The key role in this lemma was played by the existence of $T$ which is also a central element in an interaction algebra.  
We will show later in section \ref{solv} that if certain similar central elements do {\it not} appear, then there are topologically trivial (under a circuit definition) ground states.

\subsection{Other Geometries}
\label{othergeom}
Now consider this toric code Hamiltonian on a cylinder, torus, and sphere geometries.  We can take the toric code Hamiltonian on such geometries and embed it into a Hamiltonian on a finite square lattice with open boundary conditions with still a small Hilbert space dimension on each site.  Consider the case of a cylinder, first: suppose the system has $L_y$ sites in the direction with periodic boundary conditions and $L_x$ sites in the direction with open boundary conditions.
Label the site on the cylinder with coordinates $(x,y)$ with $0\leq x<L_x$ and $0\leq y <L_y$.  Then, define a system on a square lattice with length $L_x$ in the $x$-direction and $L_y/2$ in the $y$-direction with coordinates $(x,y)$ with $0\leq y \leq L_y/2$.  Let each site on the square lattice have a four-dimensional Hilbert space, with site $(x,y)$ on the square lattice containing the degrees of freedom of sites $(x,y)$ and $(x,L_y-1-y)$ on the cylinder.

This procedure can the thought of as ``squashing the cylinder flat": consider the cylinder as a two-dimensional surface embedded in three dimensions and project it onto a two-dimensional plane.  This leads to only a small increase in the Hilbert space dimension on each site.
The same kind of procedure can be applied to the torus or sphere case.

For the case of the torus Hamiltonian, there is a four-fold degenerate ground state, with different ground states having different expectation values for these central elements of the interaction algebra, and so for this Hamiltonian it might seem like considering these central elements is not important for determining whether or not there is a zero energy ground state.  Consider, however, the sphere case.  In this case, there is a unique zero energy ground state.  Suppose, however, that we make a change to the Hamiltonian on the sphere, changing the sign of the interaction term on one light plaquette.  Then, there are no zero energy ground states.  If we change the interaction sign on another light plaquette, even on a light plaquette located far away from the first plaquette with a changed sign, then there is again a zero energy ground state.  This effect is due to central elements in the interaction algebra.  Suppose we squash the sphere flat so that the north pole is at the left-most edge of our one-dimensional chain of supersites and the south pole is at the right-most edge of our chain of supersites.  If all plaquettes have the same interaction sign except possibly for those at the north and south poles, then the effective classical Hamiltonian is similar to Eq.~(\ref{effclasstc}) in that it constrains values of the central elements to be the same between neighboring supersites, but there are additional terms at the left-most and right-most supersites constraining the central elements on those sites to have a given value depending upon the sign of the interaction term on the corresponding plaquette, just as in Eq.~(\ref{effclasstc2}).

\subsection{More General Models}
In the toric code model, the central elements $O^x_C$ is a product of local operators.  Such a product of local operators is a matrix product operator with bond dimension $1$.  Using the square lattice toric code with boundaries conditions as above,the central elements $\phi_C$ of Eq.~(\ref{phiCdef}) are the operators $1+O^x_C$ which can be written as matrix product operators with bond dimension $2$.  If we consider several copies of the toric code, we can increase the bond dimension still further: consider a system with $k$-qubits on each site describing $k$ copies of the square lattice toric code.  Then the operator $\phi_C$ is $\prod_{a=1}^k (1+O^x_C(a))$, where $O^x_C(a)$ is the operator $O^x_C$ acting on the $a$-th copy.  Such an operator $\phi_C$ has bond dimension $2^k$.

However, while this toy model of several copies of the toric code shows a case where we can increase the bond dimension needed to represent $\phi_C$ as a matrix product operator, the central element $\phi_C$ is a sum of a small number (in this case, $2^k$) different central elements, each of which is a product operator (for example, one such sentral elements is $O^x_C(a)$, for any given $a$.
In contrast, if we consider more general Levin-Wen models, there may be no central elements that are product operators, and instead all the central elements may be matrix product operators (it seems likely, but we have not shown this, that a Levin-Wen model will have central elements that are product operators if and only if it is an Abelian model).

Conside a Levin-Wen model\cite{lw} on a hexagonal lattice, with the degrees of freedom on the bonds of the lattice.  This paper gives the plaquette operators $B^s_p$ as twelve-spin operators, acting on the six bonds around a given hexagon as well as the six external bonds connected to that hexagon (here, $p$ labels a given hexagon in the lattice and $s$ is a particle type).  The matrix elements of this operator $B^s_p$ are written in terms of the $F$-symbols of the theory. There is a natural generalization from this operator $B^s_p$ to an operator $B^s_l$ acting around any non-self-intersecting closed loop $l$.  Let the loop contain $n$ bonds and label the bonds along the loop as $1,2,3,....,n$ around the loop in a clockwise direction, and the external bonds connected to the loop as $e_1,e_2,e_2,...,e_n$, with bond $e_1$ in between bonds $n$ and bond $1$ and in general bond $e_a$ in between bonds $a-1$ and $a$.  The operator $B_l^s$
does not change the state of the bond $e_1,...,e_n$.
The matrix elements of $B_l$ are given by
\begin{eqnarray}
&&\langle 1',2',3',...;e_1,e_2,e_3,...| B^s_l | 1,2,3,...;e_1,e_2,e_3,...\rangle \nonumber
\\ \nonumber
&\equiv & B^{s,1',2',3',...}_{l,1,2,3,...}\\
&=&
F^{e_1n^*1}_{s^*1'n'^*} F^{e_2 1^*2}_{s^* 2' 1'^*} F^{e_3 2^* 3}_{s^* 3'2'^*} ...
\end{eqnarray}
In the above equation, the a term $e_a$ in an $F$-symbol means the state of bond $e_a$, while a term $a$ represents the initial state of bond $a$ and a term $a'$ represents the final state of bond $a$.
The last line in the above equation can be written as
\be
\prod_{a=1}^{n} F^{e_a (a-1)^* a}_{s^* a' (a-1)'^*},
\ee
with the subtraction being done mod $n$ so that $1-1=n$.

If there are $n_{types}$ particle types in the theory (plus also the identity particle), then the operator $B^s_l$ can be written as a matrix product operator of low bond dimension as follows.  Before describing how to write it as a matrix product operator, note that a {\it matrix product operator} is, by definition, an operator that acts on a one-dimensional system (a line).  The spins that we consider do not quite form a one-dimensional system, since the external bonds ``branch off" of the hexagon.  To deal with this, we will choose to combine the bonds $e_a$ and $a$ into a single degree of freedom so that we can remain in the formalism of matrix product operators.  To define a matrix product operator, we need to also introduce auxiliary degrees of freedom; let $\alpha_a$ be the auxiliary degree of freedom that couples the operator on $e_a$ and $a$ to the operator on $e_{a+1}$ and $a+1$.
Then we will show how to write
\begin{eqnarray}
\label{circleop}
B^s_l
&=& \sum_{\{\alpha\}}\Bigl(O_1^{\alpha_n \alpha_1} O_2^{\alpha_2 \alpha_1} O_3^{\alpha_3 \alpha_2} ... \Bigr)
\\ \nonumber
&=& {\rm tr}_{\{\alpha\}} \Bigl(O_1 O_2 O_3 ...\Bigr),
\end{eqnarray}
where the sum on the second line is over all possible assignments to the auxiliary degrees of freedom $\alpha$, and where $O_a^{\alpha_a \alpha_{a-1}}$ is an operator acting on $e_a$ and $a$ that depends upon the choice of the auxiliary degrees of freedom $\alpha_a$ and $\alpha_{a-1}$.
The  trace on the last line is over the auxiliary degrees of freedom $\alpha$ and the product $O_1 O_2 ...$ is a product of the matrices (viewing $O_a$ as being a matrix of operators).

We let the auxiliary degrees of freedom $\alpha_a$ be made up of two indices $\beta_a,\beta_a'$.  Each index $\beta_a,\beta_a'$ will take one of $n_{types}+1$ possible values corresponding to the different particle types, including the identity particle.  We define $O_a^{\beta_a \beta_a' \beta_{a-1} \beta_{a-1}'}$ by its matrix elements:
\begin{eqnarray}
\label{Ogiven}
&& \langle a'; e_a| O_a^{\beta_a \beta_a' \beta_{a-1} \beta_{a-1}'} | a; e_a \rangle \\ \nonumber
& = & 
\delta_{a',\beta_a'} \delta_{a,\beta_a} F^{e_a (\beta_{a-1})^* a}_{s^* a' \beta_{a-1}'^*}.
\end{eqnarray}
This may be more clearly written as
\begin{eqnarray}
&& \langle a'; e_a| O_a^{\nu \nu' \mu \mu'} | a; e_a \rangle \\ \nonumber
& = & 
\delta_{a',\nu'} \delta_{a,\nu} F^{e_a \mu^* a}_{s^* a' \mu'^*},
\end{eqnarray}
where we have set $\mu=\beta_{a-1}, \nu=\beta_a$ to reduce the number of subscripts and superscripts.

Eq.~(\ref{Ogiven}) gives the operators to write $B^s_l$ as a matrix product operator with bond dimension $(n_{types}+1)^2$.
These operators $B^s_l$ are central elements of the interaction algebra on $l$ of the term algebra.
For the doubled Fibonacci model, for example, $B^s_l$ has bond dimension $4$.

\section{Breakability}
\label{break}
In section \ref{unbreak} we used the existence of an operator $T$ which was a central element in an interaction algebra to show the unbreakability of certain operators.  
We now show a result that is in some sense a converse of this: we show that for models with certain properties of the interaction algebras, all operators which commute with the term algebra are breakable.

\begin{figure}
\includegraphics[width=1.9in]{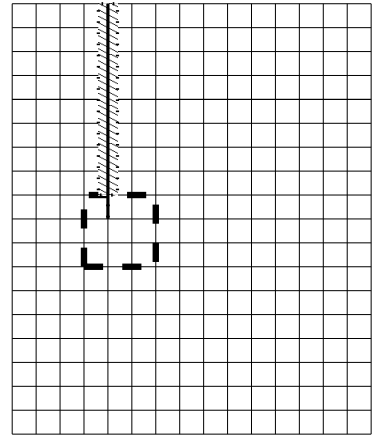}
\caption{The sites of the lattice are on the intersections of the solid horizontal and vertical lines.   The heavier dashed line represents the set $B$.  $B$ is a closed line, containing the $12$ sites around around a $4$-by-$4$ square.  The $4$ sites inside the square represent $I$ and $E$ consists of all sites not in $I$ or $B$.  The heavy solid line is $A$.  The set $X$ is $A \setminus I$, so $X$ consists of all the sites in $A$ above the line $B$ and also the intersection of $A$ and $B$.  The set $X$ has been indicated by drawing light diagonal lines around the sites in that set.}
\label{figanyon}
\end{figure}

\begin{figure}
\includegraphics[width=1.9in]{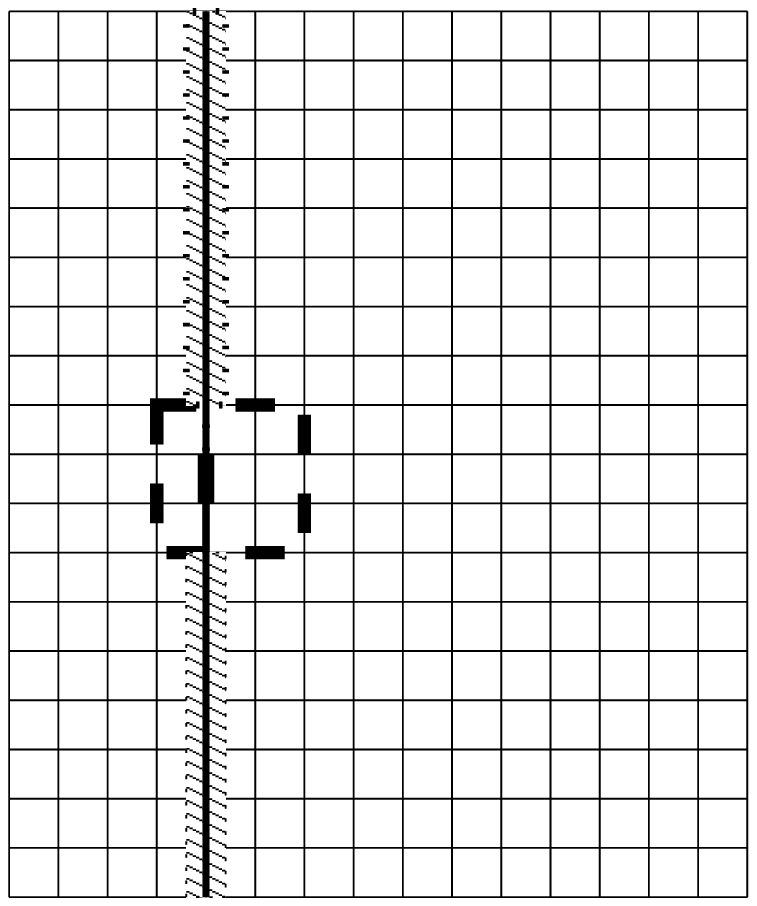}
\caption{The sites of the lattice are on the intersections of the solid horizontal and vertical lines.   The heavier dashed line represents the set $B$.  $B$ is a closed line, containing the $12$ sites around around a $4$-by-$4$ square.  The $4$ sites inside the square represent $I$ and $E$ consists of all sites not in $I$ or $B$.  The heavy solid line is $A$; one bond in this line has been draw even thicker, so that sites in $A$ above the middle of this line are in $A_1$ and sites in $A$ below the middle of this line are in $A_2$.  The sets $X,Y$ have been indicated by drawing light diagonal lines around the sites in that set, with $X$ consisting up the upper set of such points and $Y$ consisting of the lower set of such points.
}
\label{figbreak}
\end{figure}

Let $B$ be some set of sites which is the boundary of a set $S$ of sites, where the boundary of $S$ is the set of sites $i\in S$ such that $i \in Z$ for some plaquette $Z \not \subset S$.
For example, see Fig.~\ref{figanyon} or Fig.~\ref{figbreak}.  The sets $I$ and $E$ defined in the figure are the ``interior" and ``exterior" of $B$, as illustrated, with $I\cap B=E\cap B=\emptyset$; that is, $I=S\setminus B$ and $E=\Lambda \setminus S$, where $\Lambda$ is the set of sites in the whole lattice.  Note that every term $Q_Z$ is supported on $S$ or $E \cup B$.
Note that it is possible for a term $Q_Z$ to be supported on $B$ (for example, if $S$ is the set of sites in a given plaquette, so that $B=S$) and hence be supported on both $I\cup B$ and $E\cup B$.
Let ${\cal A}^I$ be the interaction algebra of $S$ on $B$ and let ${\cal A}^E$ be the interaction algebra on $B$ of the algebra generated by the $Q_Z$ for $Z\not \subset S$.

Let ${\cal H}_B$ be the Hilbert space on $B$.  
In this section we will be interested in the case that we can decompose ${\cal H}_B={\cal H}_{B\rightarrow E} \otimes {\cal H}_{B \rightarrow I}$
so that ${\cal A}^E$ acts only on ${\cal H}_{B \rightarrow E}$ and similarly ${\cal A}^I$ acts only on ${\cal H}_{B \rightarrow I}$.  This decomposition will be the condition under which we show breakability below.
Before showing breakability, we mention one case in which the absence of certain central elements implies this decomposition.
The algebras ${\cal A}^E$ and ${\cal A}^I$ commute.  So, if at least one of these algebras does not contain nontrivial central elements (note that every algebra contains the identity operator as a trivial central element; we refer to other central elements as ``nontrivial"), then this decomposition holds.

Before showing breakability of certain operators, we give a lemma which has some physical interpretation in terms of the absence of ``anyon excitations" as discussed later.
\begin{lemma}
\label{nocenbreak}
Define some set $S$ with boundary $B$ as above.  Define another set $A$ which is an arbitrary set of sites.
Define
\be
X=A\setminus I.
\ee
(We illustrate a possible choice of such sets in Fig.~\ref{figanyon}.)

Suppose ${\cal H}_B$ decomposes into ${\cal H}_{B\rightarrow E} \otimes {\cal H}_{B \rightarrow I}$ so that ${\cal A}^E$ acts only on ${\cal H}_{B \rightarrow E}$ and similarly ${\cal A}^I$ acts only on ${\cal H}_{B \rightarrow I}$.
Let $O$ be any operator supported on $A$ such that $O$ commutes with all $Q_Z$ with $Z\cap E \neq \emptyset$.
Then, we can write
\be
\label{lemres}
O=\sum_{\alpha} \Bigl( O_{X\cup B}^\alpha O^\alpha_{B\cup (A\setminus X)} \Bigr),
\ee
where the operators $O^\alpha_{X\cup B}$ are supported on $X\cup B$ and the operators $O^\alpha_{B\cup (A\setminus X)}$ are supported on $B\cup (A\setminus X)=B\cup (A\cap I)$ and where $O^\alpha_{X\cup B}$ commutes with the term algebra.
\begin{proof}
The set $A$ is a subset of the union of the disjoint sets $X\setminus B$, $B$, and $A\cap I$.
The Hilbert space ${\cal H}_B$ on $B$ can be decomposed as a product ${\cal H}_{B\rightarrow E} \otimes {\cal H}_{B\rightarrow I}$.
Thus, the Hilbert space on the support of $O$ can be decomposed as a product of two Hilbert spaces ${\cal H}_{ext} \otimes {\cal H}_{int}$, with
\be
{\cal H}_{ext}= {\cal H}_{X\setminus B}\otimes {\cal H}_{B\rightarrow E}
\ee
and
\be
{\cal H}_{int}= {\cal H}_{B\rightarrow I}\otimes {\cal H}_{A \cap I}.
\ee
Hence, we can decompose $O$ as a sum
\be
\label{hsd}
O=\sum_{\alpha} O_{ext}^\alpha O_{int}^\alpha,
\ee
where $O_{ext}$ acts on ${\cal H}_{ext}$ and $O_{int}$ acts on ${\cal H}_{int}$, and finally where we pick that
the operators ${\rm tr}\Bigl((O_{int}^\alpha)^\dagger O_{int}^{\beta}\Bigr)=0$ for $\alpha\neq \beta$ and similarly
${\rm tr}\Bigl((O_{ext}^\alpha)^\dagger O_{ext}^{\beta}\Bigr)=0$ for $\alpha \neq \beta$.  That is, we choose the operators to be orthogonal with respect to the Hilbert-Schmidt inner product.  Such a choice is possible using a singular value decomposition\cite{svdop}.

We claim that for all $\alpha$ the support of $O_{ext}^\alpha$ is a subset of $X\cup B$ and the support of $O_{int}^\alpha$ is a subset of $B\cup (A \cap I)$.  To see this, note that $O$ commutes with any operator supported on $E\setminus X$, and so using the orthogonality property of the $O_{int}^\alpha$, each operator $O_{ext}^\alpha$ commutes with any operator supported on $E\setminus X$.  Hence, $O_{ext}^\alpha$ is supported on the $X\cup B$.  The claim for $O_{int}^\alpha$ follows similarly.

Therefore, Eq.~(\ref{hsd}) gives us a decomposition of $O$ in the form of Eq.~(\ref{lemres}).  We just need to show that $O_{ext}^\alpha$ commutes with $Q_Z$ for all $Z$.  If $Z\cap I \neq \emptyset$, this follows immediately since $Q_Z$ acts on ${\cal H}_{int}$ in that case.
However, if $Z\cap I= \emptyset$, then $Q_Z$ acts only on ${\cal H}_{ext}$ and not on ${\cal H}_{int}$ and so, given that $[Q_Z,O]=0$ and that $[Q_Z,O_{int}^\alpha]=0$ and given the orthogonality property of the $O_{int}^\alpha$,  it follows that $[Q_Z,O_{ext}^\alpha]=0$ for all $\alpha$.

This completes the proof.  We note that the last paragraph is simply another application of the concept of an interaction algebra.  Let the interaction algebra of $O$ on ${\cal H}_{ext}$ be the algebra generated by the $O_{ext}^\alpha$.  Since $O$ commutes with $Q_Z$ and $Q_Z$ does not act on ${\cal H}_{int}$, the interaction algebra commutes with $Q_Z$.
\end{proof}
\end{lemma}

This last lemma \ref{nocenbreak} has a physical interpretation that there are no anyon degrees of freedom inside $I$.  For a theory like the toric code which has anyons, there are operators which are string-like which create excitations at their endpoints.  For example, consider the product of $\sigma^x_i$ over all $i$ in a column $C$ such that the vertical coordinate of $i$ is less than or equal to $y$ for some $y<L$.  This operator does not commute with the Hamiltonian, but it commutes with every term except for those terms $Q_Z$ with $Z$ acting on the endpoint of this string. Indeed, in the toric code it commutes with all but one term in the Hamiltonian, so that acting on the ground state it creates a state with energy $1$.  One can show that this operator $\sigma^x_i$ cannot be decomposed in the form of Eq.~(\ref{lemres}) for any set $B$ centered on the endpoint of the string with $B$ small enough that $B$ does not touch the boundary of the system; we omit the proof of this statement as it is very similar to the proof of lemma \ref{tcunbreak} and is based on considering commutation with a certain central element of an interaction algebra.

Similarly, one can consider the product of $\sigma^x_i$ over all $i$ in column $C$ such that the vertical coordinate of $i$ is between $x,y$ for some $1<x<y<L$.
This operator creates a pair of excitations at the ends of the string.  It commutes with all but two terms in the Hamiltonian, one near one end of the string and one near the other end.  One can similarly show that this operator cannot be decomposed into a sum of products of operators, $\sum_\alpha O_u^\alpha O_s^\alpha O_l^\alpha$,  with $O_u^\alpha$ supported near the upper end of the string, $O_l^\alpha$ supported near the lower end of the string, with the support of neither of those operators touching the boundary and with distance at least $2$ between the supports of those operators, and with $O_s^\alpha$ commuting with the Hamiltonian.  Again, the proof of this is very similar to lemma \ref{tcunbreak}.
So, physically in the toric code model one may say that these anyon excitations need to be connected by a string.

Lemma \ref{nocenbreak} shows that these properties depend upon the existence of certain central elements in the interaction algebra.  This is not surprising: we expect that if a theory has anyons then it is possible to make a measurement on $B$ which detects the existence of an anyon inside $B$, and this measurement is precisely measuring the given central elements in the interaction algebra on $B$.

Now we show the main result in this section on breakability:
\begin{theorem}
Define sets $S$ and $B$ as before.  Define a set $A$ which is an arbitrary set of sites.  Suppose that
\be
A\setminus I = X \cup Y
\ee
for two disjoint sets $X,Y$ such that for there is no plaquette $Z$ which has non-vanishing intersection with both $X$ and $Y$.  (A possible choice of such sets is
shown in Fig.~\ref{figbreak}).

Suppose ${\cal H}_B$ decomposes into ${\cal H}_{B\rightarrow E} \otimes {\cal H}_{B \rightarrow I}$ so that ${\cal A}^E$ acts only on ${\cal H}_{B \rightarrow E}$ and similarly ${\cal A}^I$ acts only on ${\cal H}_{B \rightarrow I}$.
Then, any operator $O$ supported on $A$ that commutes with the term algebra is breakable on sets $X\cup B,Y\cup B,I$.
\begin{proof}
Break $A\cap I$ into two disjoint sets, $I_1,I_2$, such that $I_1 \cap Y=I_2 \cap X=\emptyset$.  (For example, in Fig.~\ref{figbreak}, $I_1$ is the set of sites in $I$ above the middle of the heaviest solid line.)  Such sets always exist as we could pick $I_1=(A\cap I)\setminus Y$ and $A_2=(A\cap I)\setminus I_1$, though the choice in Fig.~\ref{figbreak} is a different choice of $I_1,I_2$.
Let $A_1=X\cup I_1$ and $A_2=Y\cup I_2$.
Then we can write
\be
\label{svd}
O=\sum_\alpha O_{A_1}^\alpha O_{A_2}^\alpha,
\ee
with $O_{A_1},O_{A_2}$ supported on $A_1,A_2$, respectively,  where we can choose that 
\be
\alpha \neq \beta \; \rightarrow \;
{\rm tr}\Bigl( (O_{A_1}^\alpha)^\dagger O_{A_1}^\beta \Bigr)=0
\ee
and similarly
and
\be
\label{S2obey}
\alpha \neq \beta \; \rightarrow \;
{\rm tr}\Bigl( (O_{A_2}^\alpha)^\dagger O_{A_2}^\beta \Bigr)=0.
\ee
This decomposition (\ref{svd}) can be found by a singular value decomposition.

Note that $O_{A_1}^\alpha$ fulfills the the conditions of lemma \ref{nocenbreak}.  To show this, we need to show that the commutator $[O_{A_1}^\alpha,Q_Z]=0$ for $Z$ such that $Z\cap E \neq \emptyset$.  This commutator is vanishing if $Z\cap A_1 = \emptyset$ as the supports are disjoint.  Suppose $Z\cap A_1 \neq \emptyset$.  Then, since $Z\cap I=\emptyset$,
$Z\cap X\neq \emptyset$.  Thus, $Z\cap Y=\emptyset$.  So, $[Q_Z,O_{A_2}^\beta]=0$ for all $\beta$ for such $Z$.  So, since $[O,Q_Z]=0$, using the orthogonality Eq.~(\ref{S2obey}) it follows that all the $O_{A_1}^\alpha$ commute with those $Q_Z$ also.
Similarly, the $O_{A_2}^\beta$ fulfill the conditions of the lemma also, with the set $X$ replaced with $Y$.

So, we can apply the lemma to both $O_{A_1}^\alpha$ and $O_{A_2}^\alpha$ to get
\be
\label{newalphd}
O=\sum_{\alpha} O_{X\cup B}^\alpha O_{B\cup (A_1\setminus X)}^\alpha O_{B\cup (A_2 \setminus Y)}^\alpha O_{Y\cup B}^\alpha,
\ee
where $O_{X\cup B}^\alpha$ and $O_{Y\cup B}^\alpha$ both commute with the term algebra.
Note that the index $\alpha$ in Eq.~(\ref{newalphd}) is not the same as the index $\alpha$ in Eq.~(\ref{svd}); we used Eq.~(\ref{svd}) to write $O$ as a sum of products of operators supported on $A_1$ and $A_2$ and then we used lemma \ref{nocenbreak} to write $O_{A_1}^\alpha$ as sum of products of operators on $X\cup B$ and $B\cup (A_1 \setminus X)$ and to similarly decompose $O_{A_2}^\alpha$ as another sum of products of operators on $Y\cup B$ and $B\cup (A_2\setminus Y)$, thus writing $O$ as a sum of products of operators on the four sets $X\cup B,B\cup(A_1\setminus X),B\cup(A_2 \setminus Y),Y\cup B$, as in Eq.~\ref{newalphd}.

Let $O_{I}^\alpha=O_{B\cup (A_1 \setminus X)} O_{B\cup(A_2 \setminus Y)}$, so that
\be
O=\sum_{\alpha} O_{X\cup B}^\alpha O_I^\alpha O_{Y\cup B}^\alpha.
\ee
We have thus written $O$ as a sum of products of operators with the desired support, and $O_{X\cup B}$ and $O_{Y\cup B}$ both commute with the term algebra.
So, if we could show that $O_I^\alpha$ commutes with the term algebra, then we would have shown breakability.

In fact, we cannot show that $O_I^\alpha$ commutes with the term algebra in general, but by applying a singular value decomposition again we can fix this problem.  Write the above equation as
\be
O=\sum_{\alpha} O_{X\cup Y \cup B}^\alpha O_I^\alpha,
\ee
where $O_{X\cup Y \cup B}^\alpha=O_{X\cup B}^\alpha O_{Y\cup B}^\alpha$.  By performing a singular value decomposition as before, we can write this as
\be
\label{into}
O=\sum_{\alpha} \tilde O_{X\cup Y\cup B}^\alpha \tilde O_I^\alpha,
\ee
where now the operators $\tilde O_{X\cup Y\cup B}^\alpha$ are orthogonal with respect to the Hilbert-Schmidt inner product and where $\tilde O_{X\cup Y\cup B}^\alpha$ is some linear combination of the operators $O_{X\cup Y\cup B}^\alpha$ as follows:
\be
\label{plugin}
\tilde O_{X\cup Y\cup B}^\alpha=\sum_\beta V_{\alpha\beta} O_{X\cup Y\cup B}^\beta,
\ee
for some matrix $V$ and similarly the operators $\tilde O_I^\alpha$ are linear combinations of the operators $O_I^\alpha$.  Now, using the orthogonality of $\tilde O_{X\cup Y\cup B}^\alpha$, it follows that the operators $\tilde O_I^\alpha$ commute with the term algebra.  
Plugging Eq.~(\ref{plugin}) into Eq.~(\ref{into}), we find that
\be
O=\sum_{\alpha\beta} V_{\alpha\beta} O_{X\cup B}^\beta O_{Y\cup B}^\beta O_I^\alpha,
\ee
so we have succeed in finding the desired decomposition of operator $O$; one may combine the two indices $\alpha,\beta$ into a single index to write this in the form Eq.~(\ref{breakabledef}).
\end{proof}
\end{theorem}

\section{Solvability and Trivial States Without Nontrivial Central Elements}
\label{solv}
Based on the discussion in section \ref{break}, where the absence of nontrivial central elements in a certain interaction algebra implies breakability, one might expect that 2DCLH, restricted to instances without certain nontrivial central elements, is in NP.  We now show this (see below for a description of which interaction algebras we are considering).  In the next section we consider instances with these nontrivial central elements.  In this section, where these nontrivial central elements are absent, we are able to prove more than just that the problem is in NP: we also prove triviality (under a circuit definition of topological order using ancillas as discussed below) of a zero energy ground state in theorem \ref{nocenistriv}.
This triviality then shows that if a state is nontrivial under a circuit definition (for example, a toric code ground state is nontrivial under this definition, even allowing the use of ancillas), then such nontrivial central elements must be present.

\begin{figure}
\includegraphics[width=1.9in]{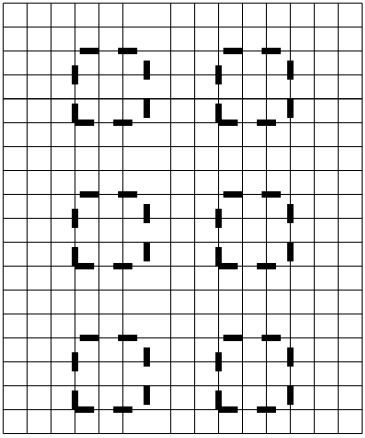}
\caption{Illustration of holes.  The sites of the lattice are on the intersections of the solid horizontal and vertical lines.  The heavier dashed lines represent the closed lines $B_i$.  There are $6$ lines, corresponds to $i=1...6$.  Each such dashed lines has been drawn around a $4$-by-$4$ square, so it contains $12$ sites.  For each line $B_i$, the $4$ sites inside the square represent $I_i$.}
\label{figholes}
\end{figure}

See Fig.~\ref{figholes}.  We have drawn several closed lines $B_i$, each line being the boundary of some set $S_i$.  
\begin{definition}
A set of {\bf holes} is a set of sets of sites, such that each set $S_i$ contains some set of four sites in a plaquette and
such that $S_i \cap S_j = \emptyset$ for $i\neq j$.
For each $i$, let $B_i$ be the boundary of $S_i$ as defined previously and $I_i$ be the interior.
Let $E_i$ be the set of sites not in $S_i$.
Let ${\cal A}_i^I$ be the interaction algebra of $S_i$ on $B_i$ and let ${\cal A}_i^E$ be the interaction algebra on $B_i$ of the algebra generated by the $Q_Z$ for $Z \not \subset S_i$. 
\end{definition}

Let ${\cal H}_{B_i}$ be the Hilbert space on $B_i$.  The algebras ${\cal A}_i^E$ and ${\cal A}_i^I$ commute.  So, if at least one of these algebras does not contain nontrivial central elements, then we can decompose 
${\cal H}_{B_i}={\cal H}_{B_i\rightarrow E_i} \otimes {\cal H}_{B_i \rightarrow I_i}$
so that ${\cal A}_i^E$ acts only on ${\cal H}_{B_i \rightarrow E_i}$ and similarly ${\cal A}_i^I$ acts only on ${\cal H}_{B_i \rightarrow I_i}$.

\begin{lemma}
Assume we have a set of holes $S_i$.
Suppose for all $i$ it is possible to decompose ${\cal H}_{B_i}$ as
${\cal H}_{B_i}={\cal H}_{B_i\rightarrow E_i} \otimes {\cal H}_{B_i \rightarrow I_i}$
so that ${\cal A}_i^E$ acts only on ${\cal H}_{B_i \rightarrow E_i}$ and similarly ${\cal A}_i^I$ acts only on ${\cal H}_{B_i \rightarrow I_i}$.
Let $H_E$ be the sum of $Q_Z$ over all $Z$ such that $Z$ is not a subset of $S_i$ for any $i$.
Let $H_I$ be the sum of $Q_Z$ over all $Z$ such that $Z$ is a subset of some $S_i$.
Then, $H$ has a zero energy ground state if and only if both $H_E$ and $H_I$ have zero energy ground states.
\begin{proof}
The ``only if" direction of the implication is obvious: since $H=H_E+H_I$ and both $H_E$ and $H_I$ are positive semi-definite, $H \geq H_E$ and $H\geq H_I$.

The ``if" direction of the implication follows from the assumption that the ${\cal H}_{B_i}$ decomposes in the given way; this implies that the Hilbert space of the whole system decomposes into ${\cal H}={\cal H}_E \otimes {\cal H}_I$ where $H_E$ acts only on ${\cal H}_E$ and $H_I$ acts only on ${\cal H}_I$ so that the spectrum of $H$ is the sum of the spectra of $H_E$ and $H_I$.  Here we have defined
\be
{\cal H}_I=\bigotimes_i \Bigl( {\cal H}_{I_i} \otimes {\cal H}_{B_i \rightarrow I_i} \Bigr).
\ee
\end{proof}
\end{lemma}

We now show that given this decomposition of ${\cal H}_{B_i}$, the problem of whether $H$ has a zero energy ground state is in NP, under some assumptions on the sets $S_i$.
The Hamiltonian $H_I$ is a sum over $i$ of Hamiltonians $H_i=\sum_{Z\subset B_i \cup I_i} Q_Z$.   
\begin{definition}
We say that a set of holes forms a {\bf square lattice of size} $O(1)$ if the following hold.
Let the centers of the holes $B_i$ form a square lattice with distance $r_{hole}$ between the centers, as shown in Fig.~\ref{figholes}, and let each hole have diameter $d_{hole}$.
Let both $d_{hole}$ and $r_{hole}$ be $O(1)$.
\end{definition}
 In Fig.~\ref{figholes}, $r_{hole}=6$.

The Hamiltonian $H_I$ is a sum over $i$ of Hamiltonians $H_i=\sum_{Z\subset B_i \cup I_i} Q_Z$.   If the holes form a square lattice of size $O(1)$, then each $H_i$ acts on a Hilbert space with dimension $D^{O(1)}$ and hence its spectrum can be determined in time $poly(D)$ by matrix diagonalization, and so whether or not $H_I$ has a zero energy state can be determined in time $N poly(D)$.
Also, in this case the Hamiltonian $H_E$ can be coarse-grained into a Hamiltonian which is a sum of commuting two-body terms, with the dimension of the Hilbert space in the coarse-grained problem being $D^{O(1)}$.  To do this coarse-graining, draw lines connecting the center of each hole to its four neighbors (above, below, left, and right of the given hole); these lines then divide the lattice of sites into different sets (deform the lines slightly so that no site is exactly on a given line); then, coarse-grain by combining all sites in one set into a single supersite, and the resulting Hamiltonian is a sum of two-body terms (here we use the assumption that every hole contains at least one plaquette).
See \cite{topotemp} for an illustration of a similar coarse-graining.
Thus, the problem of whether $H_E$ has a zero energy ground state is in NP, showing that the problem of whether $H$ has a zero energy ground state is in NP.

Note that both $H_E$ and $H_I$ have trivial ground states (trivial in the sense that it can be constructed by a local quantum circuit of depth and range $O(1)$); this result for $H_E$ follows from the fact that $H_E$ is a sum of commuting two-body terms on the supersites resulting from the coarse-graining.  We claim that 
\begin{theorem}
\label{nocenistriv}
Assume that ${\cal H}_{B_i}$ decomposes as above for a set of holes that form a square lattice of size $O(1)$.  Then,
$H$ has a trivial ground state, where we allow the use of ancillas in the construction of this ground state.  That is on each site we are allowed to tensor in additional ancilla degrees of freedom, then initialize the system to a product state, then apply the quantum circuit, so that the resulting state should be a ground state of $H \otimes I$, meaning $H$ acting on the first copy tensored with the identity on the second copy.
\begin{proof}
Define two copies of the system.  On the first copy, called the real copy, construct a trivial state that is a ground state of $H_E$ using a quantum circuit. 
Such a state follows from the coarse-graining above.  On the second copy, called the ancilla copy, construct a trivial state that is a ground state of $H_I$.  Then, apply another round of the unitary quantum circuit to swap the state on $\bigotimes_i  \Bigl( {\cal H}_{B_i \rightarrow I_i} \otimes {\cal H}_{I_i} \Bigr)$ between the two copies; this can be accomplished with a quantum circuit of depth $1$ and range $d_{hole}$; this circuit is a product of unitaries on the different $B_i \cup I_i$.  
\end{proof}
\end{theorem}
It seems possible that this construction can be modified to avoid the use of ancillas, but we do not consider this here.

We now describe an alternate way of showing that the problem of determining whether or not $H_E$ has a zero energy ground state is in NP.
This approach is based on the idea of propagation, as in Eq.~(\ref{propeq1}).  For the rest of the section, when we refer to the ``term algebra", we mean the algebra of terms of $H_E$.
The set of sites in the column $C$ which are not in any interior $I_i$ can be written as a union of several intervals of sites, which we write
$X_{1,C}, X_{2,C},...$, with no interactions coupling sites in different intervals.
Let the sets $X_{1,C},X_{2,C},...$ be sets of sites in column $C$ between the holes  as shown in the figure.  
The minimal projectors which are central elements of the interaction algebra of  the term algebra on $C$ can be written as
\be
\label{centfactors}
P^{\alpha_1,\alpha_2,...}_C=P^{\alpha_{1,C}}_{X_{1,C}} P^{\alpha_{2,C}}_{X_{2,C}} ....
\ee
wth $P^{\alpha_{1,C}}_{X_{a,C}}$ supported on $X_{a,C}$. 
So, if there exists a zero energy ground state of $H_E$, then for each column of holes, we pick a column $C$ that intersects the middle of that column of holes, and we can pick a set of $\alpha_{a,C}$ for every $a$ so that
$\prod_C \Bigl( \prod_{a} P^{\alpha_{a,C}}_{X_a,C} \Bigr)$ is consistent with $H_E$.  Note that $\prod_a P^{\alpha_{a,C}}_{X_a,C}$ is a matrix product operator with dimension $D^{O(1)}$.
So, by lemma (\ref{maskuse}) the set of these matrix product operators is a witness for the existence of a zero energy state.

\begin{figure}
\includegraphics[width=1.9in]{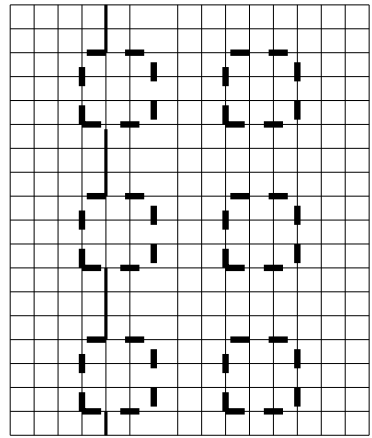}
\caption{One particular column $C$ intersecting a column of holes.  The solid lines represent the intervals $X_{a,C}$.  In this case, this is column $C=4$ and there are $4$ intervals $X_{a,C}$ for $a=1,...,4$.}
\label{figcolintersect}
\end{figure}

\section{Solvability in the General Case}
\label{solvgen}
We now consider the case in which the decomposition used above does not hold.  We actually begin with another special case: a case where the state may be topologically nontrivial but where certain density matrices are full rank and prove the existence of a witness in this case.  We then discuss the general case.  In a sense, one may regard the three cases we consider (the one in the previous section and the two in this section) as follows: the first problem considered is a problem which is essentially classical (no topological order) but where there may be ``glassiness"; the second problem considers a case where there may be topological order but where there is a certain assumption given below that one may interpret as having no glassiness; finally, we consider the case that combines both.

\subsection{Bound on Matrix Product Operator Dimension in $4$-site Model: Special Case}
In this subsection we consider a system of $4$ sites, called $1,2,3,X$.  
Consider a Hamiltonian
\be
H=Q_{12X}+Q_{2X3},
\ee
where the operators $Q_{12X}$ and $Q_{2X3}$ are positive semi-definite and Hermitian but need not be projectors.
Let $[Q_{12X},Q_{2X3}]=0$.
Let all sites have finite Hilbert space dimension, and
let the Hilbert space dimension on site $2$ be $D$ and let the Hilbert space dimension of site $3$ be $D_3$.  Note: this quantity $D_3$ will not appear in many of our bounds later, but we define this quantity because we will use it in intermediate steps of the calculation.

We define
\begin{definition}
Consider the algebra of operators on ${\cal H}_2 \otimes {\cal H}_3$ generated by the full algebra of operators on site $2$ and by the interaction algebra of $Q_{2X3}$ on site $3$.   Consider the subalgebra of this algebra consisting of operators which commute with the interaction algebra of $Q_{2X3}$ on sites $2,3$.  

Note that since this subalgebra is finite dimensional, it is a direct sum of subalgebras, each of which is isomorphic to a full matrix algebra.  We call these subalgebras ${\cal B}^\alpha_{23}$.  The Hilbert space ${\cal H}_2\otimes {\cal H}_3$ can be written as a sum of Hilbert spaces,
\be
 \bigoplus_\alpha {\cal H}_{23}^\alpha
\ee
such that the projector $P^\alpha_{23}$ onto ${\cal H}_{23}^\alpha$ is in ${\cal B}^\alpha_{23}$; indeed $P^\alpha_{23}$ is the operator that maps to the identity operator in the given full matrix algebra under the isomorphism. 
Up to a change of basis on ${\cal H}_{23}^\alpha$, the isomorphism from the full matrix algebra to ${\cal B}^\alpha_{23}$ is given by mapping a matrix in the matrix algebra to one or more copies of that matrix and the dimension of ${\cal H}_{23}^\alpha$ is an integer multiple of the dimension of the matrices.

Let the {\bf boundary algebra} be the direct sum of algebras ${\cal B}^\alpha_{23}$ over all $\alpha$ such that $P^\alpha_{23}$ does not annihilate the ground state subspace of $Q_{2X3}$.
\end{definition}

Let $|\psi^\mu_0\rangle$ be a complete basis for the ground states of $Q_{2X3}$ on sites $2,X,3$.  Then, 
since the boundary algebra commutes with $Q_{2X3}$, for every operator $O$ in the boundary algebra, we have
\be
\label{hmm}
O |\psi^\mu_0\rangle=\sum_{\mu'} X_{\mu' \mu} |\psi^{\mu'}_0\rangle,
\ee
for some matrix $X$.  We will often refer later to the matrix $X$ corresponding to an operator $O$, by which we mean the matrix $X$ that appears in Eq.~(\ref{hmm}) for the given operator $O$.
Let ${\cal X}^\alpha$ be the set of such matrices $X$ that correspond to operators $O$ in ${\cal B}^\alpha_{23}$; note that this set forms an algebra, since if operators $O,O'\in {\cal B}^\alpha_{23}$ have corresponding matrices $X,X'$, then the operator $OO'$ is in ${\cal B}^\alpha_{23}$ also and has corresponding matrix $XX'$.  Eq.~(\ref{hmm}) gives a homomorphism from ${\cal B}^\alpha_{23}$ to ${\cal X}^\alpha$.
\begin{lemma}
Suppose $P^\alpha_{23}$ does not annihilate the space of ground states of $Q_{2X3}$.  
Then, there is an isomorphism from ${\cal B}^\alpha_{23}$ to ${\cal X}^\alpha$.

Further, let the {\bf effective boundary algebra} be the direct sum of ${\cal X}^\alpha$ over $\alpha$ such that $P^\alpha_{23}$ does not annihilate the ground state subspace of $Q_{2X3}$.  Then, there is an isomorphism from the boundary algebra to the effective boundary algebra.
\begin{proof}
Suppose $\alpha$ is such that $P^\alpha_{23}$ does not annihilate the space of ground states of $Q_{2X3}$. 
The algebra ${\cal B}^\alpha_{23}$ is isomorphic to a full matrix algebra.  Suppose a given matrix $M$ in the full matrix algebra maps to a given $O$ in ${\cal B}^\alpha_{23}$ under this isomorphism.  Up to a change of basis which commutes with $Q_{2X3}$, $O$ must be some number of copies of $M$ acting on the ground state subspace and some other number of copies of $M$ acting on the rest of the space, padded with zeroes.  However, since $P^\alpha_{23}$ does not annihilate the space of ground states, there must be at least one copy of $M$ acting on the ground state subspace.  Thus, the matrix $X$ that corresponds to $O$ from Eq.~(\ref{hmm}) must be, up to change of basis, one or more copies of $M$ padded with additional zeroes.
So, this gives the desired isomorphism from ${\cal B}^\alpha_{23}$ to ${\cal X}^\alpha$.

We can similarly show the existence of an isomorphism from the boundary algebra, which is the direct sum of such ${\cal B}^\alpha$, to the effective boundary algebra, which is the direct sum of such ${\cal X}^\alpha$.
\end{proof}
\end{lemma}

We will bound the dimension of the boundary algebra under certain assumptions.  The first result, lemma \ref{lemmaspec} which is given in this subsection,will not be used later in the paper.  However, the assumptions in this lemma are applicable to many physical models as discussed later, and so this lemma is worth mentioning. Also, this construction motivates the constructions of the next two subsections.  The second result, lemma \ref{lemmaproj} which is given in the next subsection, is what we will use later.

\begin{lemma}
\label{lemmaspec}
Assume that $H$ has a unique zero energy ground state $\psi^0$.  Further, if the algebra on $2X$ generated by the interaction algebras of $Q_{12X}$ and $Q_{2X3}$ on $2X$ has central elements $P^\alpha$ projecting onto different subspaces of ${\cal H}_{2X}$, then assume that there is some given $\alpha$, which we write as $\alpha=0$, such that every state $\psi$ that is an eigenstate of $Q_{2X3}$ with zero eigenvalue also obeys $P^0 \psi = \psi$.
The subspace ${\cal H}_{2X}^0$ decomposes into two subspaces, ${\cal H}_{2X\rightarrow 1}^0 \otimes {\cal H}_{2X \rightarrow 3}^0$, with $Q_{12X}$ acting only on ${\cal H}_{2X\rightarrow 1}^0$ and $Q_{2X3}$ acting only on ${\cal H}_{2X \rightarrow 3}^0$.  We require that $Q_{2X3}$ has a unique zero energy state on ${\cal H}_{2X \rightarrow 3}^0 \otimes {\cal H}_3$. Let $K$ denote the rank of the reduced density matrix on ${\cal H}_3$ of this zero energy state.

Let $D_{BA}$ be the dimension of the boundary algebra considered as a vector space.
For such a $Q_{12X},Q_{2X3}$,
\be
D_{BA} \leq D^2 D_3/K.
\ee
In particular, if the reduced density matrix is full rank so $D_3=K$ then $D_{BA}\leq D^2$ and if $D_3/K=O(1)$ then $D_{BA}\leq O(1) D^2$.
\begin{proof}
As discussed above, the subspace ${\cal H}_{2X}^0$ decomposes into two subspaces, ${\cal H}_{2X\rightarrow 1}^0 \otimes {\cal H}_{2X \rightarrow 3}^0$, with $Q_{2X3}$ acting only on ${\cal H}_{2X \rightarrow 3}^0$.
Any ground state (in this case, this means any eigenvector with eigenvalue zero) of Hamiltonian $Q_{2X3}$ is a pure state on ${\cal H}_{2X\rightarrow 3}^0 \otimes {\cal H}_3$.  Let this pure state be $\phi$.  Introduce a basis $|\mu\rangle_{2X\rightarrow 1}$ for states on ${\cal H}_{2X\rightarrow 1}^0$.  So the ground states $|\psi_0^\mu\rangle$ above can be written as $|\mu\rangle_{2X\rightarrow 1} \otimes \phi$.

We bound the dimension of the effective boundary algebra.  Then, since there is an isomorphism from the boundary algebra (which is a direct sum of full matrix algebras) to the effective boundary algebra, this bounds $D_{BA}$.

Consider the vector space of operators acting on ${\cal H}_2 \otimes {\cal H}_3$.  This space has dimension $D^2 D_3^2$.
Every operator in the boundary algebra is in this vector space.
In fact, given any operator $O$ in the boundary algebra, and any operator $S$ acting on ${\cal H}_3$, the operator $S O$ is in the space of operators acting on ${\cal H}_2 \otimes {\cal H}_3$.  For such an operator $O$, we have that
\begin{eqnarray}
SO |\psi^\mu_0\rangle &=& S \sum_{\nu} X_{\nu \mu} |\psi^{\nu}_0\rangle \\ \nonumber
&=& \Bigl(\sum_{\nu} X_{\nu\mu}|\nu\rangle_{2X\rightarrow 1}\Bigr) \otimes \Bigl(S\phi\Bigr).
\end{eqnarray}
The above equation gives a map from the space of ground states of $Q_{2X3}$ to the space spanned by states $|\mu\rangle_{2X\rightarrow 1}$ tensored with arbitrary states on ${\cal H}_{2X\rightarrow 3}^0 \otimes {\cal H}_3$.  So, for each operator $SO$, we have some corresponding linear map between these two spaces of different dimension.  We will show that the dimension of the space of these maps is at least $D_3 K D_{BA}$.
However, this implies that the space of operators on ${\cal H}_2 \otimes {\cal H}_3$ has dimension at least $D_3 K D_{BA}$.  Hence, $D_{BA}\leq D^2 D_3/K$ as claimed.

To lower bound the dimension of the space of these maps, we define the Hilbert-Schmidt inner product of two different maps, one given by matrices $O$ (and corresponding $X$ from Eq.~(\ref{hmm})) and $S$ and the other given by $O'$ (and corresponding $X'$) and $S'$.  This inner product is
\begin{eqnarray}
\label{lminp}
&&\sum_{\mu} \langle\psi^\mu_0| O^\dagger S^\dagger S' O' |\psi^\mu_0\rangle
\\ \nonumber
&=& {\rm tr}(X^\dagger X') \langle \phi|S^\dagger S' | \phi \rangle.
\end{eqnarray}
Given an orthonormal set of matrices $X$ using the inner product ${\rm tr}(X^\dagger X')$ and another orthonormal set of matrices $S$ using the inner product $\langle \phi|S^\dagger S' | \phi \rangle$, the product of these two sets gives us an orthonormal set of linear maps using the inner product (\ref{lminp}).  Since the reduced density matrix of $\phi$ has rank $K$ on $3$, we can construct $D_3 K$ orthonormal vectors with respect to that inner product.  So, we can construct $D_3 K D_{BA}$ different linear maps which are orthonormal using inner product (\ref{lminp}).
\end{proof}
\end{lemma}

\begin{corollary}
\label{coro}
A similar bound on $D_{BA}$ holds also under the following slightly weaker assumptions.  Suppose that the interaction algebra of $Q_{2X3}$ on $3$ is not the algebra of all operators but rather a subalgebra.
Suppose the Hilbert space ${\cal H}_3$ can be decomposed into a product of two Hilbert spaces ${\cal H}_3^1 \otimes {\cal H}_3^2$, such that the interaction algebra of $Q_{2X3}$ on site $3$ only acts on the first space ${\cal H}_3^1$ in the product.  So, $Q_{2X3}$ can be written as some operator $Q_{2X3^1}$ which is supported on $2,X$ and ${\cal H}_3^1$.  We then define a new $4$-site problem, replacing site $3$ by the Hilbert space ${\cal H}_3^1$ and replacing $Q_{2X3}$ by 
$Q_{2X3^1}$.  Let ${\cal H}_3^1$ have dimension $D_{3^1}$.  For this new $4$-site problem, if it turns out that the conditions of lemma \ref{lemmaspec} hold, with the state $\phi$ having rank $K$ on ${\cal H}_3^1$ then the bound $D_{BA}\leq D^2 D_{3^1}/K$ holds.
\end{corollary}

While lemma \ref{lemmaspec} requires several assumptions about the Hamiltonian, these assumptions are in fact satisfied by many physical models after a certain grouping of sites.  We now explain this grouping (which will be used in the proof of the main theorem later), and use this grouping to show that there is a classical witness for the existence of a zero energy ground state for a certain class of 2DCLH problems (we explain which problems these are after we explain the grouping).

For each plaquette $Z$, define $v(Z)$ to be the vertical coordinate of the bottom of the plaquette.  Note that $1\leq v(Z) \leq L-1$.

\begin{lemma}
\label{almostsameas}
Consider an instance of 2DCLH.  For each column $C$, for every $i$ with $1\leq i \leq L-2$, define the following $4$-site problem, which we call the $4$-site problem centered at $i$.
Group all sites in column $C$ with vertical coordinate less than $i$ into site $1$.  Group all sites in column $C$ with vertical coordinate greater than
$i$ into site $3$.  Let site $2$ be the site in column $C$ with vertical coordinate equal to $i$.  Group all sites in columns $C-1$ and $C+1$ into site $X$.
Define $Q_{12X}$ to be the sum of $Q_Z$ over $Z$ such that $Z$ contains only sites which have been grouped into sites $1,2,X$.  Define $Q_{2X3}$ to be a weighted sum of $Q_Z$ over $Z$ such that $Z$ contains only sites which are grouped into sites $2,X,3$; these are the $Z$ such that $v(Z)>i$.
The weighted sum is a sum over such $Q_Z$, multiplying each $Q_Z$ by some generic positive scalar; the reason for this multiplication is so that every operator which commutes with $Q_{2X3}$ will commute with each such $Z$ separately; note that the zero energy states of $Q_{2X3}$ coincide with the zero energy states of an unweighted sum.

Suppose that for each $C$ and $i$, the resulting $4$-site problem obeys the conditions of lemma \ref{lemmaspec} for some given $K,D_3$ (different values of $K,D_3$ may occur for different $C,i$).  Let $(D_3/K)_{max}$ denote the maximum of $D_3/K$ over all $C,i$.

Then, for each column $C$, every central element $O_C$ of the interaction algebra of $C-1,C,C+1$ on $C$ can be written as a sum of two terms
\be
O_C=O_C^{mps}+O_C^\perp,
\ee
where $O^{mps}_C$ is a matrix product operator with dimension bounded by $D^2 (D_3/K)_{max}$ and $O_C^\perp P_{C-1,C+1}=P_{C-1,C+1}O_C^\perp=0$.
Further, if the given central element of the interaction algebra is a projector, then both $O_{mps}$ and $O^\perp$ are projectors.
\begin{proof}
Take the central element $O_C$.  For each vertical coordinate $i$, let $1_{BA}(i)$ be the unit of the boundary algebra for the $4$-site problem centered at $i$.  Then, for any $j$, $2\leq j \leq L-2$, let $O_C^{mps}(j)=1_{BA}(j) 1_{BA}(j-1) ... 1_{BA}(2) O_C 1_{BA}(2) ... 1_{BA}(j-1) 1_{BA}(j)$.
Let $O_C^{mps}=O_C^{mps}(L-2)$.

To show that this operator $O_C^{mps}$ has the desired properties, consider the $4$-site problem centered at $j$.  The operator $O_C^{mps}(j)$ commutes with $Q_{2X3}$ in the given $4$-site problem, because each of the operators $1_{BA}(k)$ for $k\leq j$ commute with the given $Q_{2X3}$.  For $k=j$, this follows by definition of the boundary algebra.  For $k<j$, it follows because $1_{BA}(k)$ commutes with the $Q_{2X3}$ in the $4$-site problem centered at $k$ and $Q_{2X3}$ in the $4$-site problem centered at $j$ appears with a generic coefficient in the weighted sum defining $Q_{2X3}$ in the $4$-site problem centered at $k$.  Note also that for all $k<j$, the operator $O_C^{mps}(k)$ commutes with $Q_{2X3}$ in the given $4$-site problem, a fact which we use in the last two paragraphs in the case $k=j-1$.

Hence, $O_C^{mps}(j)$ can be written as a sum of products of operators:
\be
O_C^{mps}(j)=\sum^\alpha O_t^\alpha O_b^\alpha,
\ee
where $O_t^\alpha$ is supported on site $1$ after grouping (supported on sites in column $C$ with vertical coordinate less than $j$) and $O_b^\alpha$ is supported on sites $2,3$ after grouping and $O_b^\alpha$ commutes with $Q_{2X3}$ in the $4$-site problem centered at $j$.  Further, because $O_C^{mps}(j)=1_{BA}(j) O_C^{mps}(j-1) 1_{BA}(j)$, the operator $O_b^\alpha$ is in the boundary algebra.  Since we have bounded the dimension of the boundary algebra by $D^2 (D_3/K)_{max}$, we have bounded the number of terms in this sum.

Write $O_C^{mps}(j)$ as a matrix product operator; we cannot bound the bond dimension on all of the bonds, but we can bound the bond dimension on bonds connecting site $i$ to $i+1$ for $i<j$.  This follows inductively since if $O_C^{mps}(j-1)$ obeys this bound, then each $O_t^\alpha$ is a matrix product operator with bounded bond dimension.
So, $O_C^{mps}$ is a matrix product operator with bond dimension at most $D^2 (D_3/K)_{max}$.

Further, since $1_{BA}(j)$ is a projector and $1_{BA}(j)$ commutes with $O_C^{mps}(j-1)$ (this follows because $1_{BA}(j)$ is in the center of the algebra of operators in sites $2,3$ that commute with $Q_{2X3}$ and $O_C^{mps}(j-1)$ commutes with $Q_{2X3}$), it follows that $O_{mps}(j)$ is a projector.

Finally, we claim that $O_C^{mps}(j)-O_C^{mps}(j-1)$ vanishes on any ground state of $Q_{2X3}$, so $O_C^\perp P_{C-1,C+1}=0$ as claimed.
To show that $O_C^{mps}(j)-O_C^{mps}(j-1)$ vanishes on ground states of $Q_{2X3}$, note that since $1_{BA}(j)$ commutes with $O_C^{mps}(j-1)$, we have $O_C^{mps}(j)-O_C^{mps}(j-1)= -O_C^{mps}(j-1) (1-1_{BA}(j))$, where $1$ is the identity operator.  The operator $1-1_{BA}(j)$ is in the center of the algebra of operators in sites $2,3$ that commute with $Q_{2X3}$, but by definition this operator vanishes on the ground state subspace.
\end{proof}
\end{lemma}
The conditions of the above lemma with $(D_3/K)_{max}=O(1)$ actually hold for many physical models, such as the toric code and Levin-Wen models with appropriate boundary conditions, such as the case in which the planar model is given by ``squashing" a cylinder into a plane (although of course for that model we know the existence of a zero energy state by other methods).
Assume a zero energy state exists a model with such a bound on $(D_3/K)_{max}$.  Here is a first attempt, which unfortunately fails, at constructing a witness to the existence of this state.
Given that a zero energy state exists, there exists a sequence of projectors $O_C$, one projector for each column $C$,
which are minimal central elements of the interaction algebra such
that ${\rm tr}(\prod_C O_C \prod_Z P_Z)>0$.  For each projector $O_C$, we write $O_C=O_C^{mps}+O_C^\perp$.  
It seems at first as if the operators $O_C^{mps}$ meet the conditions of the operators $P_C$ in lemma \ref{maskuse}.  Unfortunately, even though we know that $O_C$ commutes with $P_{C-1,C}$, we don't know that this holds for $O_C^{mps}$, preventing the use of lemma \ref{maskuse}.

There are several ways around this problem.  We will give one approach here.
The following lemma is almost the same as lemma \ref{almostsameas}; we omit the proof since it is the same except for a slightly different grouping, grouping only column $C-1$ in $X$, not $C-1$ and $C+1$ and except for the use of corollary \ref{coro} instead of lemma \ref{lemmaspec}.
Note that we could have chosen to use corollary \ref{coro} instead of lemma \ref{lemmaspec} in lemma \ref{almostsameas}; we chose not to do it there since the conditions of lemma \ref{lemmaspec} apply to many interesting problems.  However, for the next lemma, for many problems lemma \ref{lemmaspec} would not give a useful bound but corollary \ref{coro} still does; for example, for the toric code Hamiltonian, $D_{3^1}/K$ would be of order unity but $D_3/K$ would not have any useful bound with the grouping of the next lemma.
\begin{lemma}
\label{neededgroup}
Consider an instance of 2DCLH.  For each column $C$, for every $i$ with $1\leq i \leq L-2$, define the following $4$-site problem, which we call the $4$-site problem centered at $i$.
Group all sites in column $C$ with vertical coordinate less than $i$ into site $1$.  Group all sites in column $C$ with vertical coordinate greater than
$i$ into site $3$.  Let site $2$ be the site in column $C$ with vertical coordinate equal to $i$.  Group all sites in columns $C-1$ into site $X$.
Define $Q_{12X}$ to be the sum of $Q_Z$ over $Z$ such that $Z$ contains only sites which have been grouped into sites $1,2,X$.  Define $Q_{2X3}$ to be a weighted sum of $Q_Z$ over $Z$ such that $Z$ contains only sites which are grouped into sites $2,X,3$.
The weighted sum is a sum over such $Q_Z$, multipling each $Q_Z$ by some generic positive scalar; the reason for this multiplication is so that every operator which commutes with $Q_{2X3}$ will commute with each such $Z$ separately; note that the zero energy states of $Q_{2X3}$ coincide with the zero energy states of an unweighted sum.

Suppose that for each $C$ and $i$, the resulting $4$-site problem obeys the conditions of corollary \ref{coro} for some given $K,D_{3^1}$ (different values of $K,D_{3^1}$ may occur for different $C,i$).  Let $(D_{3^1}/K)_{max}$ denote the maximum of $D_{3^1}/K$ over all $C,i$.

Then, for each column $C$, every central element $O_C$ of the interaction algebra of $C-1,C$ on $C$ can be written as a sum of two terms
\be
O_C=O_C^{mps}+O_C^\perp,
\ee
where $O_{mps}$ is a matrix product operator with dimension bounded by $D^2 (D_{3^1}/K)_{max}$ and $O_C^\perp P_{C-1,C}=P_{C-1,C}O_C^\perp=0$.
Further, if the given central element of the interaction algebra is a projector, then both $O_{mps}$ and $O^\perp$ are projectors.
\end{lemma}.

Then, it follows that
\begin{theorem}
\label{toponoglass}
Consider an instance of 2DCLH which has a zero energy ground state.  
Suppose the conditions of lemma \ref{neededgroup} hold, for $(D_3^1/K)_{max} \leq poly(N)$.  Then, there exists a witness to the existence of that ground state which can be efficiently verified.
\begin{proof}
Given that a zero energy state exists, there exists a sequence of projectors $O_C$, one projector for each column $C$,
which are minimal central elements of the interaction algebra such
that ${\rm tr}(\prod_C O_C \prod_Z P_Z)>0$.
For each projector $O_C$, write $O_C=O_C^{mps}+O_C^\perp$, with $O_C^{mps},O_C^\perp$ as in lemma \ref{neededgroup}.  The projector
 $O_C^{mps}$ commutes with $P_{C-1,C}$ (this follows from the fact that $O_C$ commutes with $P_{C-1,C}$ and that $O_C^\perp P_{C-1,C}=0$ for this approach).  Also, $O_C^{mps}$ is a mask as follows from the fact that $O_C$ is a mask and that $O_C^\perp P_{C-1,C}=0$.  So, by lemma \ref{maskuse}, the set of projectors $O_C$ form a witness to the existence of a zero energy state. 
\end{proof}
\end{theorem}
Note that the conditions of this theorem hold for many interesting Hamiltonians such as Levin-Wen and toric code (though of course for those models we know the existence of zero energy ground states by other means).

\subsection{Bound on Matrix Product Operator Dimension in $4$-site Model: More General Case}
We now return to the $4$-site model, but consider different assumptions.

\begin{lemma}
\label{lemmaproj}
Consider the $4$-site model.  Suppose that there is no projector $P$ in the boundary algebra such that the partial trace ${\rm tr}_2(P^0_{2X3} P)$ is not full rank on $3$, where $P^0_{2X3}$ is the projector onto the zero energy subspace of $Q_{2X3}$.

For such a $Q_{12X},Q_{2X3}$, for each $\alpha$ such that $P^\alpha$ does not annihilate the ground state subspace of $Q_{2X3}$,
the dimension of the boundary algebra, considered as a vector space, is bounded by $D_{BA} \leq D^4$.
\begin{proof}
We bound the dimension of the effective boundary algebra, as in the proof of lemma \ref{lemmaspec}.

Consider the vector space of operators acting on ${\cal H}_2 \otimes {\cal H}_3$.  This space has dimension $D^2 D_3^2$.
Every operator in the boundary algebra is in this vector space.
In fact, given any operator $O$ in the boundary algebra, and any operator $S$ acting on ${\cal H}_3$, the operator $S O$ is in the space of operators acting on ${\cal H}_2 \otimes {\cal H}_3^0$.  For such an operator $O$, we have that
\be
\label{regard}
SO |\psi^\mu_0\rangle=S \sum_{\nu} X_{\nu\mu} |\psi^{\nu}_0\rangle.
\ee
The algebra of matrices $X$ is a direct sum of one or more algebras $X^\alpha$ which are isomorphic to full matrix algebras ${\cal M}^\alpha$.    Consider the algebra ${\cal M}^\alpha$ for some given index $\alpha$ and let it have dimension $D_\alpha$.
For each such algebera ${\cal M}^\alpha$, consider it as a vector space and consider the subspace of the vector space corresponding to matrices which are non-zero only in the first row (of course, this choice is somewhat arbitrary, as we could make an arbitrary change of basis).  Let $V^\alpha$ be the vector space of matrices in $X^\alpha$ that maps to this given vector space under the isomorphism.
Choose a basis for $V^\alpha$ consisting of matrices that map to a matrix with a $1$ in the $i$-th column of the first row and a zero everywhere else, for each $1\leq i \leq D_\alpha$.  Call the matrices in this basis $X(i,\alpha)$.

We define the Hilbert-Schmidt inner product of two different maps, one given by matrices $O$ (and corresponding $X$) and $S$ and the other given by $O'$ (and corresponding $X'$) and $S'$.  This inner product is
\begin{eqnarray}
\label{defining}
&&\sum_{\mu} \langle\psi^\mu_0|O^\dagger S^\dagger S' O'|\psi^\mu_0\rangle \\ \nonumber
&=&\sum_{\mu\nu\rho} (X^{\dagger})_{\mu \nu}   X'_{\rho \mu}   \langle\psi^\nu_0| S^\dagger S' |\psi^\rho_0\rangle.
\end{eqnarray}
Note that if $X=X(i,\alpha)$ and $X'=X(j,\beta)$, then this inner product is zero unless $i=j$ and $\alpha=\beta$.  Further, for each $i,\alpha$, we can set $X=X'=X(i,\alpha)$ and use Eq.~(\ref{defining}) to define an inner product for the $S$, with $D_3^2$ orthonormal basis vectors.
This is the point at which we use the assumption that the reduced density matrix on site $3$ of $P^0_{2X3} P$ for every projector $P$ in the boundary algebra is full rank, since the inner product for $S,S^\dagger$ is ${\rm tr}_3(S S^\dagger \rho_3)$, for some density matrix $\rho_3$ which is a reduced density matrix of such a $P^0_{2X3} P$.

So, this gives us $\sum_\alpha D_\alpha D_3^2$ different  choices of $X,S$ which are orthonormal using inner product (\ref{defining}).  So, $\sum_\alpha D_\alpha \leq D^2$.  This implies that $\sum_\alpha D_\alpha^2 \leq D^4$.
\end{proof}
\end{lemma}

We have the following generalization of this lemma, which is slightly different to corollary \ref{coro}, using instead a direct sum decomposition:
\begin{lemma}
\label{coro2}
A similar bound on $D_{BA}$ holds also under the following slightly weaker assumptions.  Suppose that the interaction algebra of $Q_{2X3}$ on $3$ commutes with some projector $P_3$ on ${\cal H}_3$.  Let the dimension of the range of $P_3$ be $D_{3^1}$.
Suppose that all ground states of $Q_{2X3}$ are in the range of $P_3$.  Construct the $4$-site problem as above. For this new $4$-site problem, if it turns out that the conditions of lemma \ref{lemmaproj} hold, with the state $\phi$ having rank $K$ on ${\cal H}_3$ then the bound $D_{BA}\leq D^2 D_{3^1}/K$ holds.
\begin{proof}
Since the interaction algebra commutes with $P_3$, we can write it as a direct sum of an algebra which does not annihilate $P_3$ and an algebra that does annihilate $P_3$.  Every operator in the second algebra annihilates every vector in the ground state of $Q_{2X3}$.  The dimension of the algebra that does not annihilate $P_3$ is at most $D_{3^1}^2$.  We then use the same proof as in lemma \ref{lemmaproj}.
\end{proof}
\end{lemma}

\subsection{Reduced Dimensionality in General Case}
The above lemma leads to the following result in the case of a general instance of 2DCLH.  Consider a general instance for which a zero energy state exists.  Then, there exists a sequence of projectors $O_C$, one projector for each column $C$,
which are minimal central elements of the interaction algebra such
that ${\rm tr}(\prod_C O_C \prod_Z P_Z)>0$.  Suppose that these projectors cannot be written as low bond dimension matrix product operators.  
Using lemma \ref{lemmaproj} or \ref{coro2}, a result analogous to lemma \ref{neededgroup} can be proven, showing the existence of  a decomposition $O_C=O_C^\perp+O_C^{mps}$, assuming that the conditions of lemma \ref{lemmaproj} or \ref{coro2} are met for every grouping for some $D_{3^1}/K$ which is at most $poly(N)$ (we abbreviate this by simply saying  that the conditions of the lemma ``are met" below, rather than every time referring to the ratio $D_{3^1}/K$).
So, if $O_C$ is not a low bond dimension matrix product operator, either the conditions of lemma \ref{coro2} are not met for some grouping, or $1_{BA}$ is not equal to the identity operator for some grouping.  We claim that in either case, we can construct a projector $\Pi$ supported on the set of sites in column $C$ with vertical coordinate greater than or equal to $j$, for some $j$ such that $\Pi$ commutes with all the $Q_Z$ except possibly for those $Q_Z$ with $v(Z)=j$, and such that $\Pi$ is consistent with the Hamiltonian, and such that $\Pi$ is not full rank.

To prove this claim, suppose first that $1_{BA}$ is not the identity operator for some grouping.  Then, simply take $\Pi=1_{BA}$.  Instead, suppose that the conditions of \ref{coro2} are not met for some grouping centered at $i$.  Then, there is some projector $R$ in the boundary algebra such that its reduced density matrix on sites $i+1,i+2,...$ is not full rank.  Suppose that there is a ground state $\psi$ such that $\langle \psi, R \psi \rangle \neq 0$.  In this case, we will take $\Pi$ to be the projector onto the range of this reduced density matrix.  We now show that $\Pi$ has the desired properties.   Let $X$ be the plaquette in columns $C-1,C$ with $v(Z)=i$ and let $Y$ be the plaquette in columns $C,C+1$ with $v(Z)=i$.  Then, $Q_X Q_Y R \psi$ has non-zero norm, as may be verified by noting that $|Q_X Q_Y R \psi|\geq |\langle \psi,Q_X Q_Y R \psi \rangle|=|\langle \psi, R \psi \rangle|$.  Further, $Q_X Q_Y R \psi$ is a ground state, as $R$ commutes with all $Q_Z$ except for $Z=X$ and $Z=Y$, and $\Pi Q_X Q_Y R \psi=Q_X Q_Y R \psi$, so $\Pi$ is consistent with the Hamiltonian.  The commutation requirements on $\Pi$ hold also.

Suppose instead that no such ground state $\psi$ exists.  In this case, we take $\Pi=1-R$.  Thus, in either case, such a projector exists.

Using these projectors $\Pi$ leads to the following idea for a proof in the general case.  We treat even and odd columns differently, constructing the groupings and projectors $O_C$ only for even $C$.
We define a new problem, $H'=\sum_Z Q'_Z$, where in a deviation from our previous notation each projector $Q'_Z$ is supported on the set of all sites in the same column as a site in $Z$ with a vertical coordinate greater than or equal to $v(Z)-1$ (that is, the sites in $Z$ as well as all sites ``below" them).
Given two projectors $A,B$, define $A\cap B$ to project onto the intersection of the range of $A$ and $B$, and define $A \cup B$ to equal $1-(1-A)\cap (1-B)$.  So, we follow the following iterative procedure.  Start by defining $Q'_Z$ to be cup over all $Q_Y$ such that $Y$ is in the same two columns as $Z$ and $v(Y)\geq v(Z)$.  Then, support of the $Q'_Z$ is indeed as we claimed.
  Then, if the operators $O_C$ are not low bond dimension matrix product operators, construct a projector $\Pi$ as discussed above, and replace $Q'_Z=Q'_Z \cup (1-\Pi)$ for all $Z$ such that $Z$ intersects column $C$ and such that the support of $\Pi$ does not include any sites above $Z$.  
This new problem still has a zero energy ground state, because $\Pi$ was consistent with $H$.
We repeat this construction again and again for different groupings until no more such $\Pi$ can be found.

Now, if this new Hamiltonian $H'$ were still defined by a set of commuting projectors, this would allow us to find a witness for $H$, because using results above we could show that central elements of $H'$ are low bond dimension matrix product operators.  (The fact that the support of $H$ has been increased will not affect our use of lemma \ref{lemmaproj} since we really only used properties of $Q_{2X3}$ there; the $4$-site problem for $H'$ now has terms $Q_{12X3}$ and $Q_{2X3}$ but we only use properties of the latter term).
However, the new Hamiltonian may not be a set of commuting projectors: the projectors $Q'_X$ and $Q'_Y$ need not commute if $v(X)=v(Y)$ and $X$ and $Y$ intersect on an even column.  This may occur because $\Pi$ may not commute with $Q_Z$.
We will explain elsewhere how to deal with this situation.

\section{Discussion}
We have proven the existence of classical witness to the existence of a zero energy ground state for certain instances of 2DCLH; a general proof will be given elsewhere.  Other results in sections \ref{example},\ref{break},\ref{solv} help characterize possible central elements in the interaction algebra in these systems.

One interesting feature is that we naturally write certain central elements in the interaction algebra as matrix product operators.  We have also shown how to do this specifically for certain central elements in Levin-Wen models.  This representation as a matrix product operator may have 
useful practical applications in attempts to engineer a system that realizes a Levin-Wen model, perhaps using Josephson junctions or cold atoms.  At first glance, realizing a Levin-Wen model seems completely impractical, since one needs to realize twelve spin operators.
However, the ability to represent these operators at matrix product operators may help lead to a practical realization by interacting given degrees of freedom with a low dimensional auxiliary system; this leads to an interesting question of physically implementing a matrix product operator with bounded bond dimension.  We do not consider this question further.

The bound on the dimension of the boundary algebra may be interesting for use elsewhere.  This bound can be interpreted as a bound on the number of particle types possible in a topological quantum field theory described by a lattice model.  After all, to create a pair of anyons in such theories, one acts with a stringlike operator that commutes with the Hamiltonian everywhere except at its ends.  This stringlike operator does not annihilate the ground state.  Assuming this operator is in the algebra generated by the interaction algebra and by the algebra of operators at its endpoints, this stringlike operator is then in the boundary algebra.

We relate the existence of certain central elements to the circuit definition of topological order in section \ref{solv}, showing
triviality of the ground state of $H$ in section \ref{solv} in the absence of certain central elements.

Of course, a natural question is what happens in three dimensions.  Is the problem of  whether a sum of commuting projectors with local interactions on a three dimensional cubic lattice has a zero energy ground state a problem that is in NP?  The discovery of codes like Haah's code\cite{haah,haah2} in three dimension casts some doubt on whether this can be true.  These codes show that there might be some very ``wild" possibilities in three dimensions which make the decision problem much harder.

{\it Acknowledgments---} I thank the Banff International Research Station (BIRS) for hospitality during the workshop on ``Operator Structures in Quantum Information Theory" while some of this work was completed.


\begin{thebibliography}{99}
\bibitem{mpoexp} To briefly summarize the definitions of \cite{mpo,mpo2} and to fix notation: given $L$ sites on a line, labelling the sites by integers $0,1,...,L-1$, a matrix product operator representation of an operator $O$ is given by writing
$O=\sum_{\alpha\beta\gamma....\tau} O_0^{\alpha} O_1^{\alpha\beta} O_2^{\beta\gamma} ... O_{L-1}^\tau$ where $\alpha,\beta,\gamma,...$ are discrete indices, and $O_0^{\alpha}$ is an operator supported on site $0$ (there is one such operator for each choice of $\alpha$), $O_1^{\alpha\beta}$ is an operator supported on site $1$ (there is one such operator for each choice of the pair of indices $\alpha,\beta$), and so on.  The number of values that index $\alpha$ can take is called the bond dimension on the bond connecting sites $0$ and $1$, the number of values that $\beta$ can take is called the bond dimension on the bond connecting sites $1$ and $2$.  The bond dimension of the matrix product operator representation is the maximum over all bonds of the bond dimension on each bond.
Every operator $O$ can be written as a matrix product operator but for an arbitrary $O$ the bond dimension needed might be exponentially large; we are interested in this paper in the case where the matrix product operator representation has a small bond dimension.  Note that the same operator $O$ might have different matrix product operator representations with different bond dimensions, and strictly speaking we should talk about the bond dimension of a matrix product operator representation rather than a matrix product operator.  However, we will often speak of the bond dimension of a matrix product operator meaning by this the bond dimension of a particular representation of that operator.  Also, we should note that the above definition applies to a line, while if the sites are on a circle then a slightly different definition is more natural: $O=\sum_{\alpha\beta\gamma....\tau\mu} O_0^{\mu\alpha} O_1^{\alpha\beta} O_2^{\beta\gamma} ... O_{L-1}^{\tau\mu}$, and we use this in Eq.~\ref{circleop}.
\bibitem{svdop} Using the Hilbert-Schmidt inner product, introduce an orthonormal basis of operators $X_{ext}^\alpha$ on ${\cal H}_{ext}$ and another orthonormal basis of operators $X_{int}^\beta$ on ${\cal H}_{int}$.  Then, $O$ can be written as a sum of products of these operators as $O=\sum_{\alpha\beta} A_{\alpha\beta} X_{int}^\alpha X_{ext}^\beta$, where $A_{\alpha\beta}$ is a complex scalar depending upon $\alpha,\beta$.  Then, Eq.~(\ref{hsd}) follows from a singular value decomposition of the matrix $A_{\alpha\beta}$.  This decomposition is used in the theory of matrix product operators.
\end{thebibliography}
\end{document}